\documentstyle[epsf]{elsart}

\newcommand{\be}{\begin{equation}}
\newcommand{\ee}{\end{equation}}
\newcommand{\bea}{\begin{eqnarray}}
\newcommand{\eea}{\end{eqnarray}}

\begin{document}
\bibliographystyle{prsty}

\begin{frontmatter}

\title{Sideband Instabilities and Defects of Quasipatterns}

\author{Blas Echebarria\thanksref{Co}}
\author{\and Hermann Riecke\thanksref{HR}}

\address{
Department of Engineering Sciences and Applied Mathematics,
Northwestern University, 2145 Sheridan Rd, Evanston, IL, 60208, USA
}

\thanks[Co]{Corresponding author. Tel.: (617) 373-2924, Fax.: 
(617) 373-2943, e-mail: blas@presto.physics.neu.edu. Current address:
Physics Department, Northeastern University, Boston, MA 02115.}
\thanks[HR]{E-mail: h-riecke@northwestern.edu}
\begin{abstract}

Quasipatterns have been found in dissipative systems
ranging from Faraday waves in vertically vibrated fluid
layers to nonlinear
optics. We describe the dynamics of
octagonal, decagonal and dodecagonal quasipatterns by means
of coupled Ginzburg-Landau equations and study their
stability to sideband perturbations analytically using long-wave
equations as well as by direct numerical simulation. Of particular interest
is the influence of the phason modes, which are associated
with the quasiperiodicity, on the stability of the patterns.
In  the dodecagonal case, in contrast to the octagonal and the decagonal case, 
the phase modes and the phason modes
decouple and there are parameter regimes in which the
quasipattern first becomes unstable with respect to phason
modes rather than phase modes. We also discuss the different
types of defects that can arise in each kind of quasipattern
as well as their dynamics and interactions. Particularly
interesting is the decagonal quasipattern, which allows two
different types of defects. Their mutual interaction can be
extremely weak even at small distances.

\end{abstract}

\end{frontmatter}

\section{Introduction}

Since the discovery of quasicrystals in 1984 \cite{ShBl84}, much attention has
been paid to the properties of these materials. In contrast to perfect crystals
they lack periodicity, but preserve long-range orientational order. Due to this lack of 
periodicity quasicrystals may have non-crystallographic rotational symmetry and, in fact,
materials have been found with five, eight, ten or twelve-fold rotational axes.

In dissipative systems the possibility of quasiperiodic structures, or
quasipatterns, was also suggested some time ago \cite{MaNe89}. Since then they have
been observed in Faraday waves with one- and two-frequency forcing \cite{ChAl92,EdFa93} and in nonlinear optics \cite{PaRa95}. Marangoni 
convection with a deformable interface has also been suggested to support quasipatterns \cite{GoNe95a}, although they have not been observed so far.

In Faraday waves a great variety of patterns has been observed. They include
superlattices, rhombic states, oscillons, as well as quasipatterns
\cite{KuPi98,ArFi98,ArFi00}. In order for superlattices and quasipatterns to be stable
the mutual suppression of plane-wave modes of different orientation has to be sufficiently 
weak. This can be due to a few, somewhat different mechanisms.
 When forced with two frequencies, the system can exhibit simultaneously  
instabilities at two different wavelengths and the quasipatterns and superlattices 
appear near this bicritical point.  
The angle between the wavevectors of the destabilizing modes depends on the ratio of the
two wavelengths. It determines whether the resulting patterns corresponds to a superlattice
or a quasipattern. A 
mechanism involving two length scales was suggested some time ago for 
quasicrystals \cite{MeTr85} and studied in dissipative systems by means of
a modified Swift-Hohenberg equation with two marginal modes \cite{LiPe97}. The second 
length scale  need not be associated with a proper instability. It can be sufficient that 
one of the modes is weakly damped. The interaction of the unstable mode with this 
damped mode  can then lead to a reduction in the mutual suppression of modes 
subtending a certain angle \cite{SiTo00}. Alternatively, the interaction with the damped mode
can strongly enhance the saturating self-coupling term which then effectively 
leads to a reduction of the competition over a wide range of angles \cite{Mu94,ZhVi96}.
This favors patterns with a large number of modes. Calculations of the 
coefficients of the amplitude equations for Faraday waves suggest that this is 
indeed the case \cite{ZhVi97,ChVi99}.
 
In nonlinear optics the optical field in a nonlinear cavity can be rotated in
order to obtain a structure with the desired number of modes \cite{PaRa95}.
In these systems it is also possible to arrange patterns with two different
spatial scales, leading to complex quasicrystaline structures 
\cite{ReRa96,PaRe97,PiRu99}.

The relative stability of perfect quasipatterns has already been addressed \cite{MaNe89}. In the present paper we are interested in the stability of
$n$-fold quasipatterns to sideband perturbations, in particular to slowly
varying modulations. We will assume that the physical 
fields can be expanded as a sum of Fourier modes rotated by $2\pi /n$ 
relative to each other (similar to the density wave picture of quasicrystals), with slowly varying amplitudes in space and time. Using symmetry arguments 
we determine a set of coupled Ginzburg-Landau equations for these amplitudes. 
Due to the lack of periodicity, a rigorous derivation from the basic equations
using a center manifold reduction has not been possible so far. However, 
we will consider the Ginzburg-Landau equations as phenomenological models and aim to extract 
conclusions that can be verified experimentally. 

As is well known in the study of quasicrystals, the long-wave dynamics of the
system is governed by two types of modes: phonons and phasons. The former ones
are marginal modes arising from spatial translation symmetries. The phasons, on the 
other hand, are additional marginal modes characteristic of quasicrystals and
appear because of the quasiperiodic nature of these structures. As we will 
see later, in octagonal and dodecagonal quasipatterns they have a very simple
geometrical interpretation, since these quasipatterns can be viewed as two 
superimposed square (hexagonal) lattices rotated by $2\pi /8$ ($2\pi /12$) .
The phason modes correspond to {\it relative} translations between the two
lattices. It is worth mentioning that when the angle between the two lattices
satisfies certain conditions, the whole structure becomes periodic with a
wavelength larger than that of the individual lattices: this is known as a
superlattice \cite{DiSi97}. In this case, extra resonance conditions among the modes are
satisfied \cite{SiPr98}, and the phason modes become damped and can, in principle, be
eliminated in a long-wave analysis.

The paper is organized as follows. In the next three sections 
we study octagonal, decagonal and dodecagonal quasipatterns arising from
steady bifurcations. In each case
we start with the amplitude equations and study various simple solutions and 
their relative stability. Our main focus is the long-wave dynamics of the
quasiperiodic solutions. We derive coupled equations for the phonon- and
phason-type modes, obtaining the diffusive counterpart to the elastic 
equations for a quasicrystal 
\cite{HuWa00}. We then calculate the long-wave stability limits for sideband perturbations and 
investigate numerically the behavior arising from the instabilities. In 
section 5 we discuss the different types 
of defects that appear in each kind of quasipattern and their dynamics and 
interactions. Conclusions are presented in section 6.

\section{Octagonal Quasipatterns}

\subsection{Ginzburg-Landau Equations}

\begin{figure}
\centerline{
\epsfxsize=4cm\epsfbox{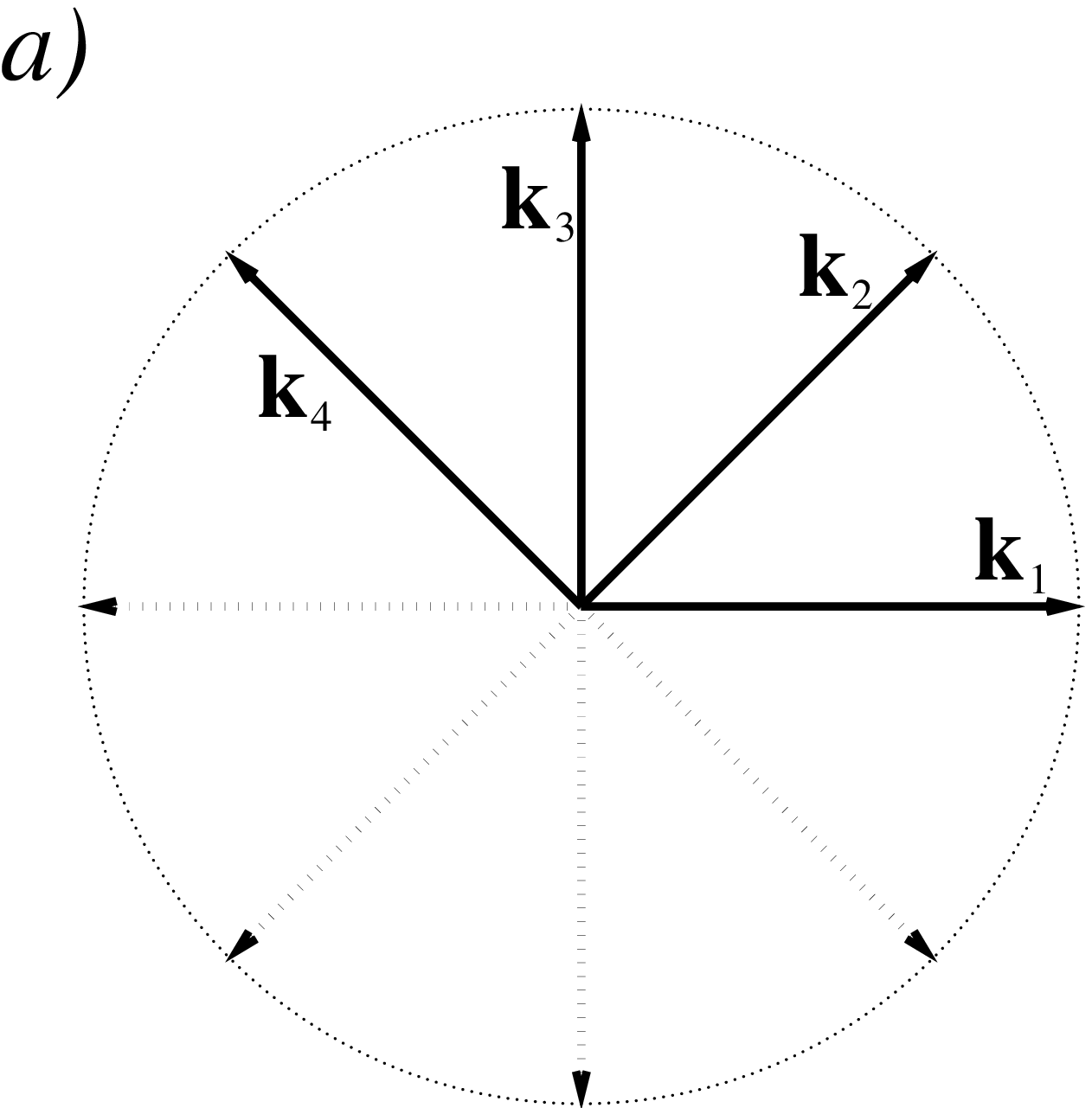}\hspace{0.5cm}
\epsfxsize=4cm\epsfbox{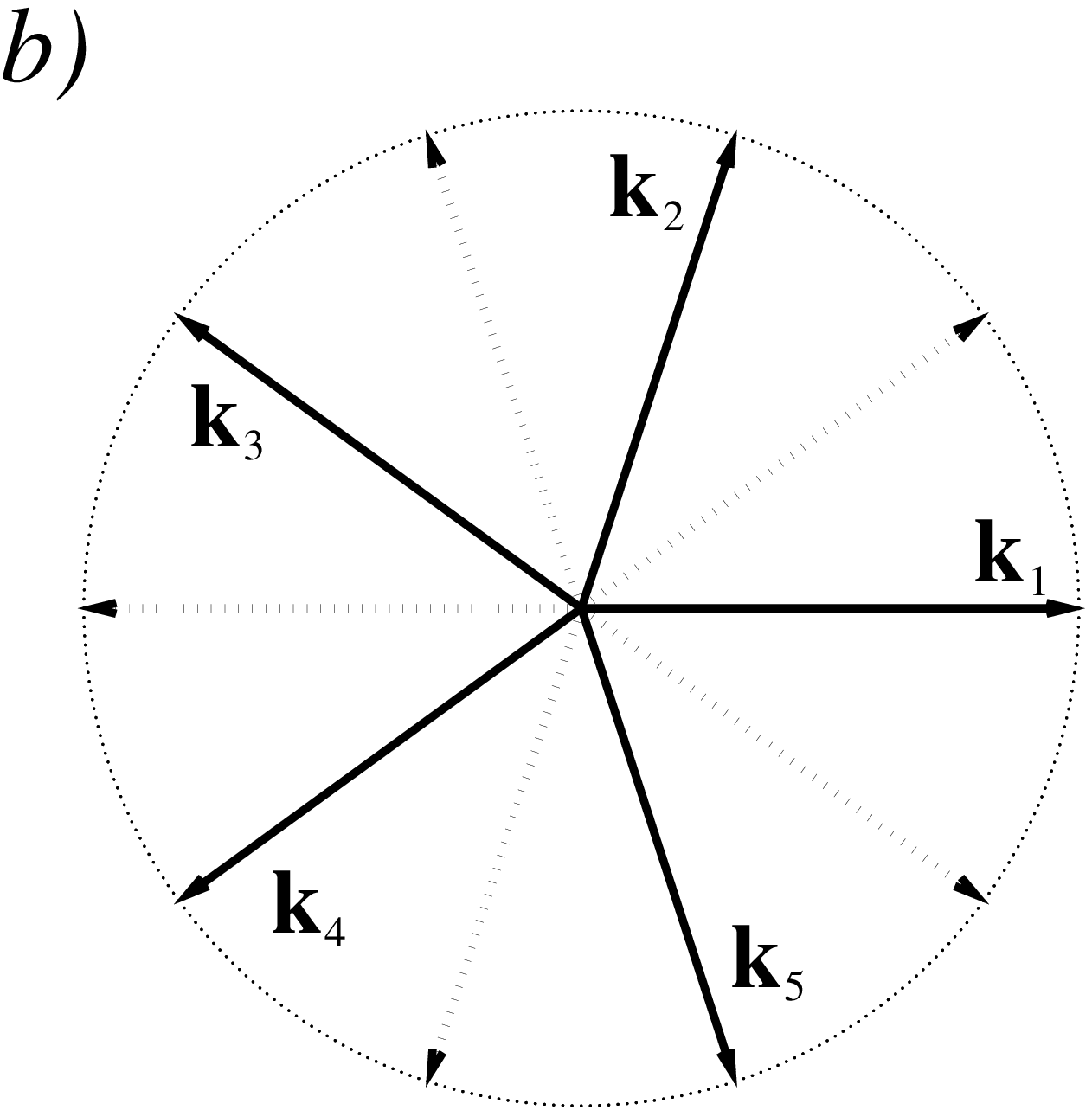}\hspace{0.5cm}
\epsfxsize=4cm\epsfbox{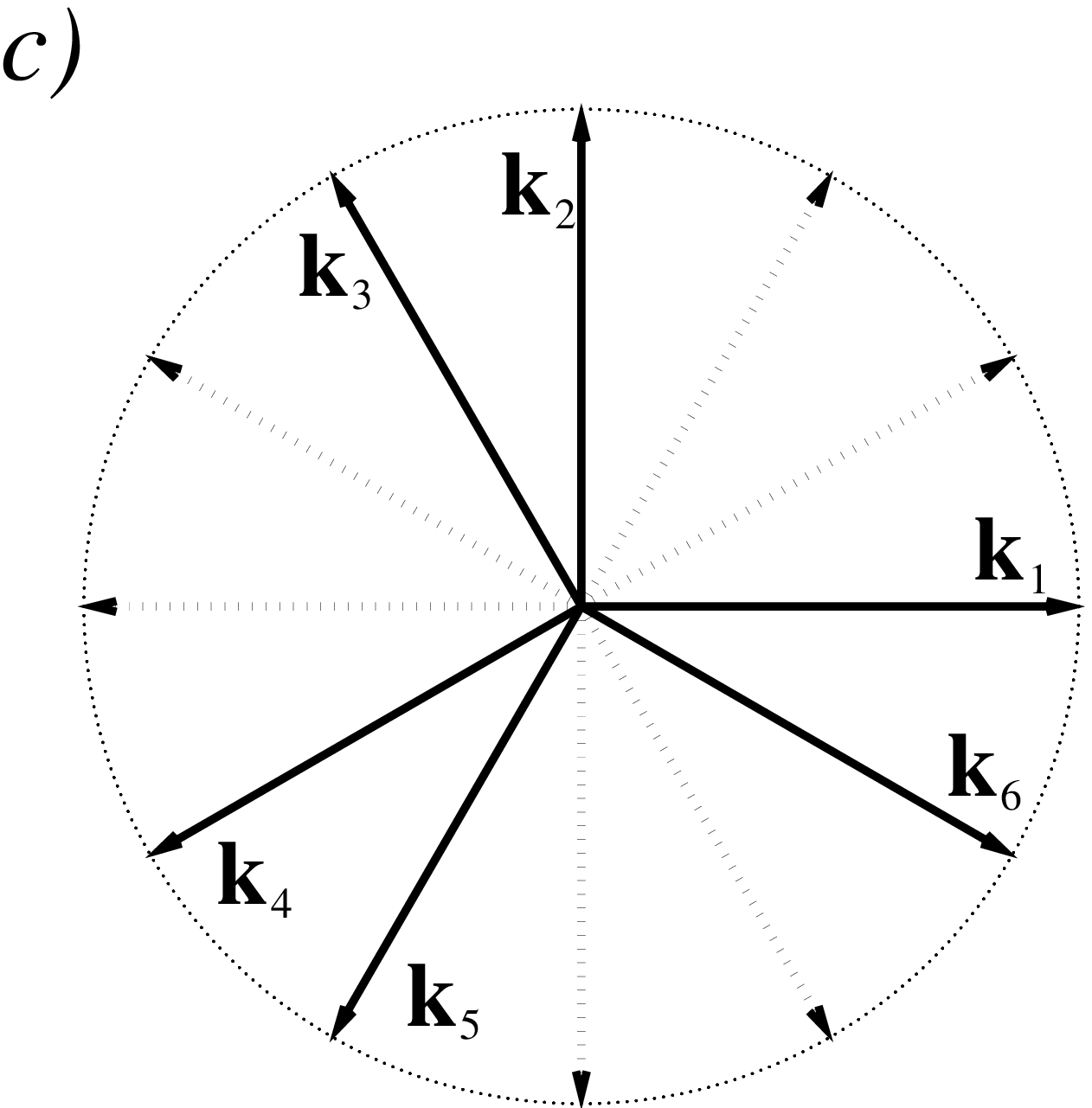}}
\caption{Fourier modes composing the a) octagonal, b) decagonal and
c) dodecagonal quasipatterns. Solid vectors denote the modes described by
Eqs. (\ref{eq.GLn8}), (\ref{eq.GLn10}), and (\ref{eq.GLn12.1},
\ref{eq.GLn12.2}), respectively.
\label{fig.star}}
\end{figure}

We consider a quasipattern composed of four modes, $\psi = \eta \sum_{n=1}^{4}
A_n({\bf x},t) e^{i{\bf k}_n \cdot {\bf \hat{x}}}+c.c.+h.o.t$, where $\eta \ll 1$
and the amplitudes 
$A_i({\bf x},t)$ satisfy the equations (after rescaling):
\begin{equation}
\partial_t A_i = \mu A_i + (\hat{\bf n}_i \cdot \tilde{\nabla})^2 A_i 
- A_i \left [|A_i|^2 + \nu |A_{i+2}|^2 + 
\gamma ( |A_{i+1}|^2 + |A_{i+3}|^2 ) \right ], 
\label{eq.GLn8}
\end{equation}
with $\mu$ the control parameter and $\hat{\bf n}_i$ the unit vector in the 
direction of the wavenumber ${\bf k}_i$ (see Fig. \ref{fig.star}a). 
 The amplitudes $A_i$ depend on slow scales $x=\eta \hat{x}$
and $t=\eta^2 \hat{t}$. Here and
in the following the indices are repeated cyclicly with period 4. Thus, $A_5$
corresponds to $A_1$, etc. The
coefficients $\nu$ and $\gamma$ measure the interaction between modes that subtend an
angle of $\pi/2$ and $\pi/4$, respectively. 

Eqs. (\ref{eq.GLn8}) have gradient structure
\begin{equation}
\partial_t A_i = -\frac{\partial {\cal F}}{\partial \overline{A}_i},
\end{equation}
and can be derived from the Lyapunov functional
\begin{eqnarray}
{\cal F}&\equiv&\int \int dxdy F =\int \int dxdy \sum_{i=1}^{4} \left [ 
-\mu |A_i|^2 
+ |(\hat{\bf n}_i \cdot \tilde{\nabla})^2 A_i|^2 + \frac{1}{2}|A_i|^4 
\right .\nonumber \\
&& \left . + \gamma |A_i|^2 |A_{i+1}|^2  + \frac{\nu}{2}
|A_i|^2 |A_{i+2}|^2 \right ].  
\end{eqnarray}
Therefore, the dynamics of the system are relaxational.

In order to calculate the different solutions we write 
$A_j = a_j e^{i\phi_j}$, with $a_j$ real. Splitting into real and
imaginary parts yields
\begin{eqnarray}
\partial_t a_i & = & \mu a_i  - a_i[a_i^2 + \nu a_{i+2}^2  
+ \gamma ( a_{i+1}^2 + a_{i+3}^2 ) ],\\ 
\partial_t \phi_i & = & 0.
\end{eqnarray}
The amplitudes can be considered to be real ($\phi_1=\cdots =\phi_4=0$). There are five different kinds of simple solutions \cite{MaNe89} (see Fig. \ref{fig.solsn8}):

\begin{figure}
\centerline{
\epsfxsize=4.5cm\epsfbox{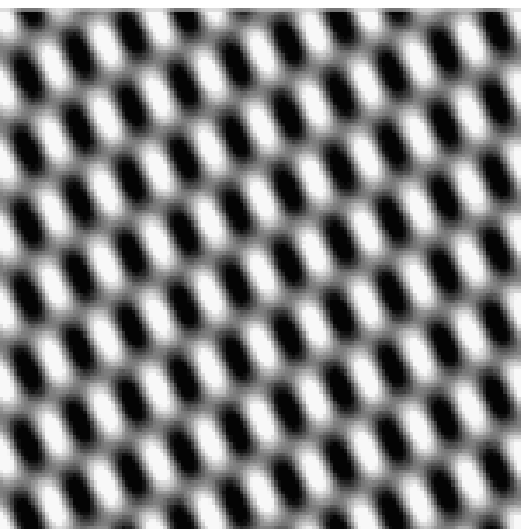}\hspace{0.5cm}
\epsfxsize=4.5cm\epsfbox{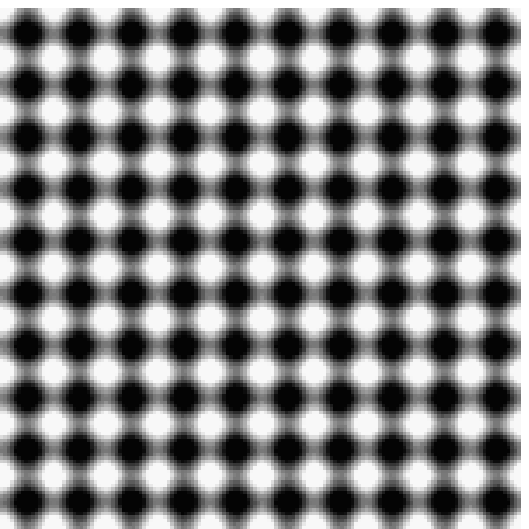}
}
\vspace{0.5cm}
\centerline{
\epsfxsize=4.5cm\epsfbox{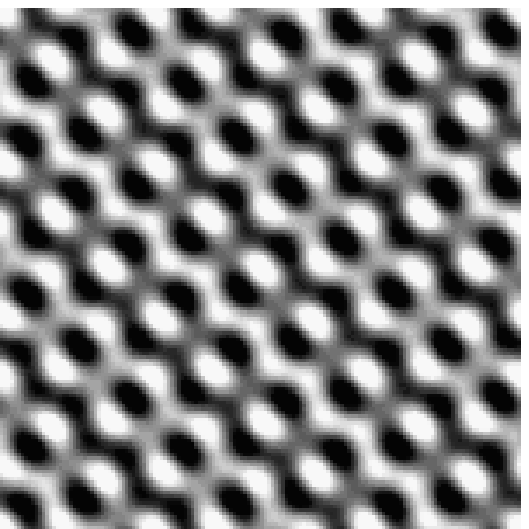}\hspace{0.5cm}
\epsfxsize=4.5cm\epsfbox{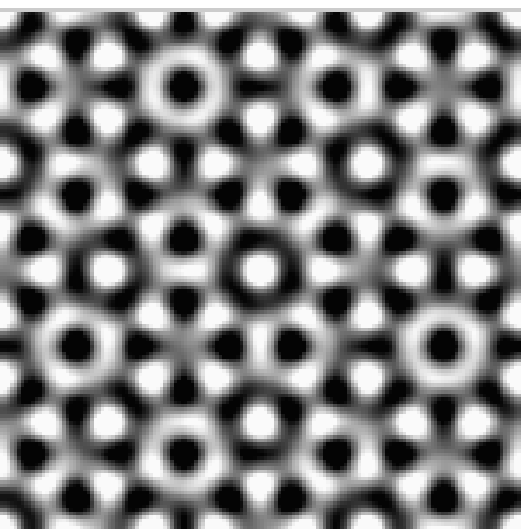}
}
\caption{Octagonal case. Different types of solutions of Eqs. (\ref{eq.GLn8}), obtained
by reconstructing the physical field $\psi$ from the amplitudes $A_i$. a) Rectangles, b) squares, c) 1D quasipattern and d) octagonal quasipattern. \label{fig.solsn8}}
\end{figure}

\begin{enumerate}

\item Rolls: $a_2=a_3=a_4=0$ and $a_1=\sqrt{\mu}$. The value of the 
Lyapunov functional for rolls is: $F_{Roll}=-\mu^2/2$.\\

\item Squares: $a_2=a_4=0$ and $a_1=a_3=\sqrt{\mu/(\nu+1)}$,  $F_{Sq}=-\mu^2/(\nu+1)$.\\

\item Rectangles: $a_3=a_4=0$ and $a_1=a_2=\sqrt{\mu/(\gamma+1)}$, $F_{Rect}=-\mu^2/(\gamma+1)$.\\

\item One-dimensional quasipattern: $a_4=0$ and
\begin{equation}
a_2=\sqrt{\frac{\mu (1+\nu - 2\gamma)}{1+\nu  -2\gamma^2}},\;\;
a_1=a_3=\sqrt{\frac{\mu (1-\gamma)}{1+\nu -2\gamma^2}}.
\end{equation}
\begin{equation}
\Rightarrow\;\;\;F_{1DQ}=-\frac{\mu^2}{2}\frac{3+\nu-4\gamma}{1+\nu-2\gamma^2}.
\end{equation}
It is quasiperiodic in the direction of ${\bf k}_2$ and periodic along 
${\bf k}_4$. As is typical for solutions with submaximal isotropy,
this solution exists only over a limited range of the nonlinear coefficients,
 $(1+\nu-2\gamma)(1-\gamma)>0$. It is always unstable \cite{MaNe89}.\\

\item Quasipattern: $a_1=\cdots =a_4=R$, with $R$ satisfying,
\begin{equation}
R=\sqrt{\frac{\mu}{1+\nu+2\gamma}}\;\;\;\Rightarrow\;\;\;F_Q=-\frac{2\mu^2}{1+\nu+2\gamma}. \label{eq.Qsoln8}
\end{equation}

\end{enumerate}

The conditions for linear stability of these solutions are given by
\begin{enumerate}

\item  Rolls: $\nu>1$, $\gamma>1$.\\

\item Squares: $\gamma>(1+\nu)/2$, $-1<\nu<1$.\\

\item Rectangles: $\nu>1$, $-1<\gamma<1$.\\

\item Quasipattern: $-(1+\nu)/2 < \gamma < (1+\nu)/2$, $-1<\nu < 1$.

\end{enumerate}

\subsection{Longwave analysis}

We assume that the conditions for the two-dimensional 
quasipattern to be stable with respect 
to homogeneous perturbations are
satisfied and study its stability to sideband perturbations. In particular,
we study the stability of the pattern as a function of its wavenumber.
A perfect quasipattern with wavenumber ${\bf k}$ slightly different 
from critical is given by $A_i=Re^{iq\hat{\bf n}_i\cdot {\bf x}}$, with
$R=|A_1|=|A_2|=|A_3|=|A_4|=\sqrt{(\mu - q^2)/(1+\nu +2\gamma)}$ and 
$\eta q=|{\bf k}- {\bf k}_c |$. We expand around this solution, 
$A_i = (R+r_i)e^{i(q\hat{\bf n}_i \cdot {\bf x} + \phi_i)}$. Considering first 
only homogeneous perturbations ($r_i=r_i(t)$, $\phi_i=\phi_i(t)$) and separating real and imaginary parts, the linearized equations for the perturbations are:
\begin{eqnarray}
\partial_t r_i &=& - 2 R^2 r_i - 2\nu R^2 r_{i+2}
- 2\gamma R^2 (r_{i+1} + r_{i+3}), \label{eq.n8linr}\\
\partial_t \phi_i & = & 0. \label{eq.n8linp}
\end{eqnarray}

For the amplitude perturbations there are three eigenvalues, two corresponding 
to one-dimensional eigenspaces, $\sigma_{\hat{r}_1} = - 2 R^2 (1+ \nu + 
2\gamma)$, for $v_{\hat{r}_1}=[r_1=1,r_2=1,r_3=1,r_4=1]/2$, and $\sigma_{\hat{r}_2}= 
-2R^2 (1+\nu - 2\gamma)$ for $v_{\hat{r}_2}=[-1,1,-1,1]/2$. The other 
eigenvalue $\sigma_{\hat{r}_3,\hat{r}_4}=-2R^2(1-\nu)$ corresponds to a two dimensional eigenspace spanned by $v_{\hat{r}_3}=[-1,-1,1,1]/2$, $v_{\hat{r}_4}=[-1,1,1,-1]/2$.

The four phases are marginal modes. Physically it is more convenient to
rearrange them into phonon and phason modes. The first two are 
related to translations in space and can be written as
\begin{eqnarray} 
&&u_{\phi_x}^{T}=[\phi_1=1, \phi_2=\cos(2\pi/8), \phi_3=\cos(4\pi/8), \phi_4=\cos(6\pi/8)], \\
&&u_{\phi_y}^{T}=[0, \sin(2\pi/8), \sin(4\pi/8), \sin(6\pi/8)]. 
\end{eqnarray}
There still remains a two-dimensional subspace orthogonal to this one. An
orthonormal basis for it is given by 
\begin{eqnarray}
&&u_{\varphi_1}^{T}=[1,-\cos(2\pi/8), \cos(4\pi/8), -\cos(6\pi/8)],\\ 
&&u_{\varphi_2}^{T} =[0, -\sin(2\pi/8),\sin(4\pi/8), -\sin(6\pi/8)]. 
\end{eqnarray}
This choice of the phason modes corresponds to relative translations (in the
x- and y-directions, respectively) between the two square
lattices that compose the quasipattern. It is worth mentioning that the phason
modes do not transform as a vector. In fact,
under rotations by an angle $\theta$ the transformation of 
the phason field $\tilde{\varphi}=(\varphi_1,\varphi_2)$ corresponds to that of a regular
vector for a rotation by an angle $5\theta$ (cf. (\ref{eq.transphonon}.\ref{eq.transphason})).

When spatial modulations are included the perturbation equations become
\begin{eqnarray}
\partial_t r_i &=& -2qR(\hat{\bf n}_i \cdot \tilde{\nabla})\phi_i 
+ (\hat{\bf n}_i \cdot \tilde{\nabla})^2 r_i  \label{eq.n8linr.mod} \\
&&- 2 R^2 r_i - 2\nu R^2 r_{i+2}
- 2\gamma R^2 (r_{i+1} + r_{i+3}),  \nonumber \\
\partial_t \phi_i & = & (\hat{\bf n}_i \cdot \tilde{\nabla})^2 \phi_i + \frac{2q}{R}(\hat{\bf n}_i \cdot \tilde{\nabla})r_i. \label{eq.n8linp.mod}
\end{eqnarray}
Now the phase modes are no longer marginal, but exhibit diffusive
dynamics. In order to study these long-wave dynamics we define a small parameter 
$\epsilon$ and introduce slow time and space scales: 
$T =\epsilon t$, ${\bf X} = \epsilon^{1/2}{\bf x}$. The amplitude and phase 
perturbations are expanded as:
\begin{eqnarray}
r_i({\bf X},T) &=& \epsilon \sum_{j=1}^{4} \hat{r}^i_{j}({\bf X},T) v^i_{\hat{r}_j}\\
\phi_i({\bf X},T) &=& \epsilon^{1/2} [ \phi_x({\bf X},T) u^i_{\phi_x} + 
\phi_y({\bf X},T) u^i_{\phi_y}, \nonumber \\
&& + \varphi_1({\bf X},T) u^i_{\varphi_1} + \varphi_2({\bf X},T) u^i_{\varphi_2} ]. 
\end{eqnarray}

Inserting the expansion into Eqs. (\ref{eq.n8linr.mod}), (\ref{eq.n8linp.mod}), we 
obtain at order $\epsilon^{1/2}$  
the eigenvalues for the homogeneous perturbations, at order $\epsilon$ a 
relation between the stable and marginal modes of the type $\hat{r}_i=
f_i (\nabla \phi_x,\nabla \phi_y,\nabla \varphi_1,\nabla \varphi_2)$, and at order 
$\epsilon^{3/2}$ a
solvability condition for the marginal modes. This gives us the slow,  
longwave dynamics. In component form the final equations are:
\begin{eqnarray}
\partial_T \phi_x&=&D_1\nabla^2 \phi_x + (D_2 - D_1) \partial_X (\nabla 
\cdot \vec{\phi})  \label{eq.phase.n8.1}\\ 
&&+ D_3 (\partial^2_{X}\varphi_1 - 2\partial^2_{XY} \varphi_2 - \partial^2_{Y} \varphi_1 ), \nonumber \\
\partial_t \phi_y&=&D_1\nabla^2 \phi_y + (D_2 - D_1) \partial_Y (\nabla 
\cdot \vec{\phi}) \\
&&+ D_3 (-\partial^2_{X}\varphi_2 - 2\partial^2_{XY} \varphi_1 + 
\partial^2_{Y} \varphi_2 ), \nonumber \\
\partial_t \varphi_1 &=& \frac{1}{2} \left [ (D_4 + D_5) \nabla^2 \varphi_1 +
(D_4 - D_5)(\partial_X^2 \varphi_2 - \partial_Y^2 \varphi_2 -2\partial_{XY}^2
\varphi_1)\right ] \\
&&+ D_3 (\partial^2_{X}\phi_x - 2\partial^2_{XY} \phi_y - \partial^2_{Y} \phi_x ), 
\nonumber \\
\partial_t \varphi_2 &=&  \frac{1}{2} \left [ (D_4 + D_5) \nabla^2 \varphi_2 +
(D_4 - D_5)(\partial_X^2 \varphi_1 - \partial_Y^2 \varphi_1 + 2\partial_{XY}^2
\varphi_2)\right ] \label{eq.phase.n8.4} \\
&&+ D_3 (-\partial^2_{X}\phi_y - 2\partial^2_{XY} \phi_x + \partial^2_{Y} \phi_y ), \nonumber
\end{eqnarray}
where the values of the coefficients are given by
\begin{eqnarray}
&&D_1 = \frac{1}{4} - \frac{q^2}{u},\label{eq.n8D1}\\
&&D_2 = \frac{3}{4} - \frac{q^2}{u} - \frac{2q^2}{v_1}=D_1 + D'_2,\\
&&D_3 = D_1,\\
&&D_4 = D_1,\\
&&D_5= \frac{3}{4} - \frac{q^2}{u} - \frac{2q^2}{v_2}=D_1 + D'_5 ,\label{eq.n8D5}
\end{eqnarray}
with $v_1=2R^2(1+\nu+2\gamma)$, $v_2=2R^2(1+\nu  - 2\gamma)$, $u=2R^2 (1-\nu)$, 
and $R$ given by Eq. (\ref{eq.Qsoln8}) with $\mu$ replaced by $\mu-q^2$. 
Eqs. (\ref{eq.phase.n8.1})-(\ref{eq.phase.n8.4}) are the diffusive analogue of the 
elastic equations for octagonal quasicrystals, and their form 
can be deduced directly by means of symmetry arguments.

A more compact notation, in which the symmetries become more evident, can be 
achieved by writing the phonon and phason modes as complex fields. 
Let $\phi =\phi_x + \rm{i} \phi_y$, $\varphi = \varphi_1 + \rm{i} \varphi_2$ 
and
$\nabla=\partial_X + \rm{i}\partial_Y$. Then the former expressions become:
\begin{eqnarray}
&&\partial_t \phi = D_1 |\nabla|^2 \phi + (D_2 - D_1) \nabla 
(\overline{\nabla} \phi + \nabla \overline{\phi}) + D_3  e^{\rm{i}\alpha} 
\overline{\nabla}^2 \overline{\varphi}, \\
&&\partial_t \varphi = \frac{1}{2}(D_4 + D_5) |\nabla|^2 \varphi + \frac{\rm{i}}{2} (D_4 - D_5) \nabla^2 \overline{\varphi} + D_3 e^{\rm{i}\alpha} 
\overline{\nabla}^2 \overline{\phi}
\end{eqnarray}
The angle $\alpha$ depends on the basis that is chosen for the phonon and phason
modes. Since the specific choice does not have any physical relevance,
$\alpha$ does not
appear in the dispersion relation. In our case, $\alpha= 0$. 
Now it is easy to see that these equations are
invariant under rotations of $n\pi/4$. In fact, under such a rotation $\phi \rightarrow e^{n\pi\rm{i}/4}
\phi$, $\nabla \rightarrow e^{n\pi\rm{i}/4} \nabla$, $\varphi \rightarrow 
e^{5n\pi\rm{i}/4} \varphi$, and consequently
\begin{equation}
\overline{\nabla}^2 \overline{\phi}\;\; \rightarrow \;\; e^{-2n\pi\rm{i}/4}
\overline{\nabla}^2 e^{-n\pi\rm{i}/4} \overline{\phi}=e^{-3n\pi\rm{i}/4}
\overline{\nabla}^2 \overline{\phi},\label{eq.transphonon}
\end{equation}

\begin{equation}
\overline{\nabla}^2 \overline{\varphi}\;\; \rightarrow \;\; e^{-2n\pi\rm{i}/4}
\overline{\nabla}^2 e^{-5n\pi\rm{i}/4} \overline{\varphi}=e^{n\pi\rm{i}/4}
\overline{\nabla}^2 \overline{\varphi}.\label{eq.transphason}
\end{equation}
Thus, while $\overline{\nabla}^2 \overline{\phi}$ transforms as $\varphi$,  
$\overline{\nabla}^2 \overline{\varphi}$ transforms as $\phi$. 

In order to calculate the stability boundaries, normal modes must be 
considered: $\phi=\phi_0 e^{i{\bf Q}\cdot {\bf x}}$, 
$\varphi=\varphi_0 e^{i{\bf Q}\cdot {\bf x}}$. From Eqs. (\ref{eq.phase.n8.1})-(\ref{eq.phase.n8.4}), it can be shown that the stability
curves are given by the expression:
\begin{equation}
2D_1 D_2 D_4 D_5 = D^2_3 [(D_1 + D_2)(D_4 + D_5) - 2 D_3^2 ],
\end{equation}
or, using Eqs. (\ref{eq.n8D1})-(\ref{eq.n8D5}),
\begin{equation}
D^2_1 D'_2 D'_5 =0.
\end{equation}
Thus, two eigenvalues become zero when $D_1=0$ and the other two when 
$D'_2=0$ and $D'_5=0$, respectively. From this, the stability curves are 
given by:
\begin{eqnarray}
&&D_1=0\;\;\Rightarrow\;\;\mu_1 = \frac{3 + \nu + 4\gamma}{1-\nu} q^2,\\
&&D'_2=0\;\;\Rightarrow\;\;\mu_2 = 3q^2,\\
&&D'_5=0\;\;\Rightarrow\;\;\mu_5= \frac{3+3\nu+2\gamma}{1+\nu-2\gamma} q^2.
\end{eqnarray}

\begin{figure}
\centerline{
\epsfxsize=7cm\epsfbox{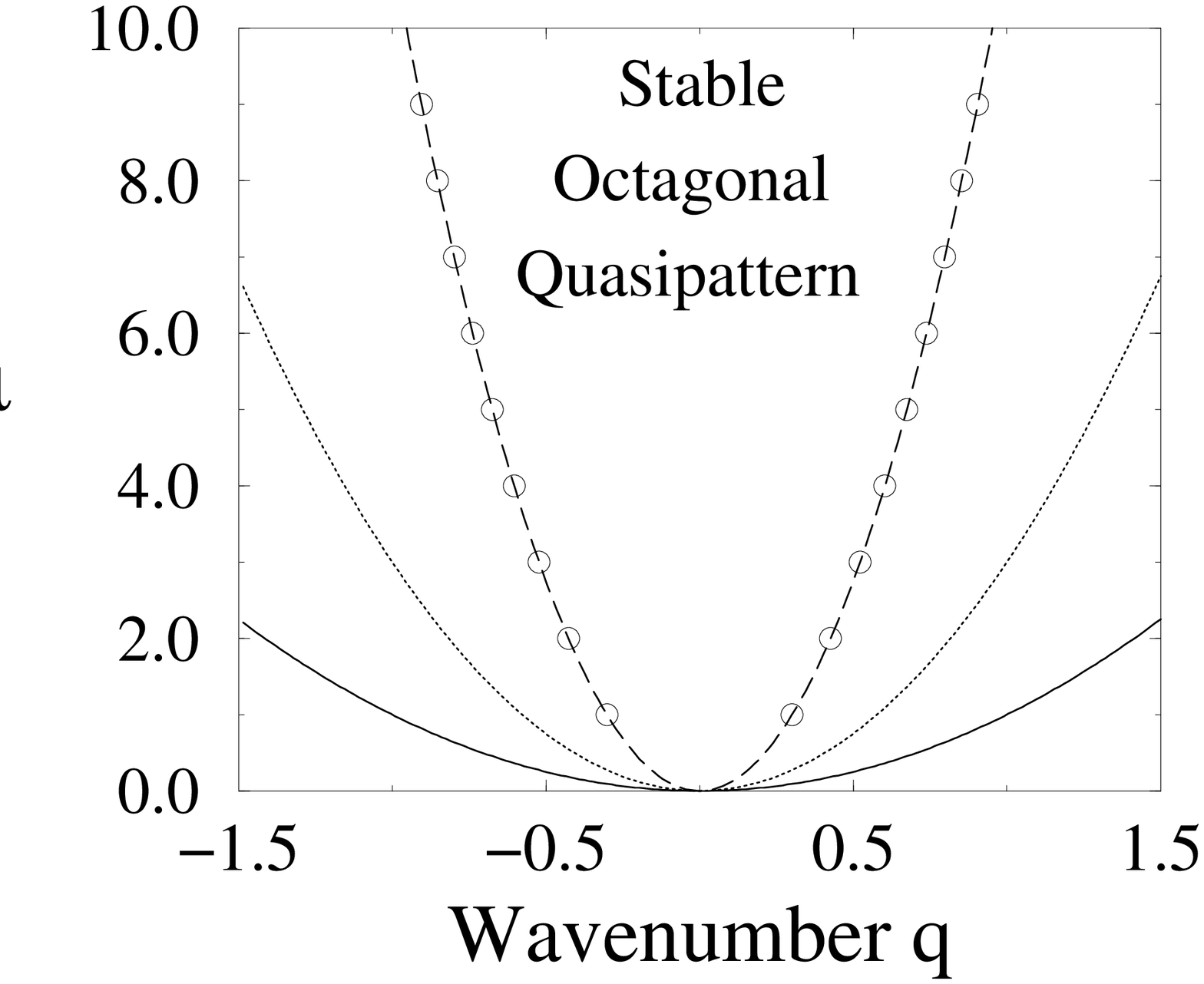}\hspace{0.5cm}
\epsfxsize=7cm\epsfbox{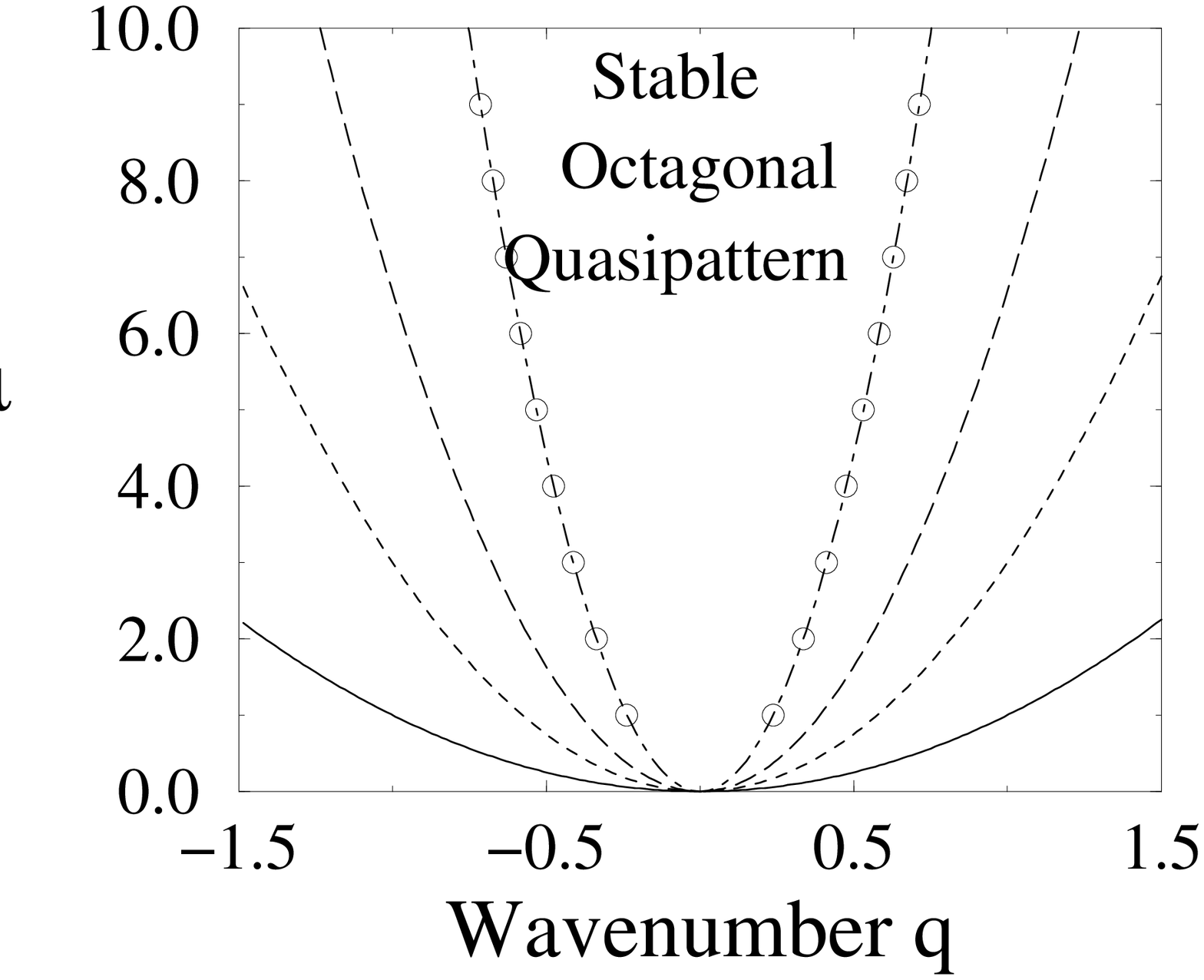}
}
\caption{Stability diagrams for a) $\nu=\gamma=0.5$, b) $\nu=0.7$, 
$\gamma=0.4$. The dotted, dashed, and dot-dashed lines correspond to
$D'_2=0$, $D'_5=0$ and $D_1=0$, respectively. The circles are obtained solving
the full dispersion relation associated with Eqs. (\ref{eq.n8linr.mod}), 
(\ref{eq.n8linp.mod}).
\label{fig.stab8}}
\end{figure} 

It is interesting to note that the instability corresponding to $D'_2=0$ 
is given by the usual value for the Eckhaus curve. Typical stability diagrams are shown in Fig. \ref{fig.stab8}. When $\nu=\gamma$
the problem is degenerate and $\mu_1 = \mu_5$ (Fig. \ref{fig.stab8}a). Which
instability occurs first depends on the values of the coefficients $\nu$ and
$\gamma$. For $\gamma>0$, $\nu<\gamma$; $\nu+\gamma <0$, $\gamma<0$; and $\nu+\gamma>0$, 
$\nu>\gamma$ the first instability is that corresponding to
$D'_5$, $D'_2$, and $D_1$, respectively. The third 
case is shown in Fig. \ref{fig.stab8}b. The symbols denote the results of a stability analysis 
using the full dispersion relation without the long-wave approximation. They show that 
the relevant instabilities are long-wave.

In general, the eigenvalues depend on the angle of the perturbation. 
When $\theta=\arctan(Q_y/Q_x)=n\pi/4$ there is an
eigenvalue that is always marginal. (From Eqs. (\ref{eq.n8linr.mod}), (\ref{eq.n8linp.mod}) it is easy to see that the growth rate of a perturbation to the phase $\phi_i$ in a direction perpendicular to $\hat{\bf n}_i$, is zero). This means that the perturbations
perpendicular to each set of rolls evolve on a still slower time scale that is
not captured with Eqs. (\ref{eq.GLn8}). 
This situation is equivalent to that of the zig-zag instability of 
rolls or squares \cite{NeWh69,Ho93}.  In order to resolve 
this degeneracy we must take Newell-Whitehead-Segel derivative terms in Eqs. (\ref{eq.GLn8}), using the replacement 
\begin{equation}
(\hat{\bf n}_i\cdot \nabla)^2\;\; \rightarrow \;\; [(\hat{\bf n}_i\cdot \nabla) - \rm{i}\delta \nabla^2]^2,
\end{equation}
where $\delta$ is a small coefficient, whose size depends on the 
distance from threshold, $\delta =O(\eta)$.

\begin{figure}
\centerline{
\epsfxsize=4.5cm\epsfbox{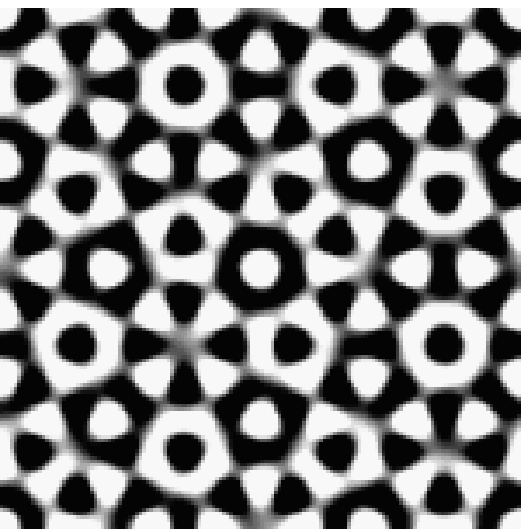}\hspace{0.5cm}
\epsfxsize=4.5cm\epsfbox{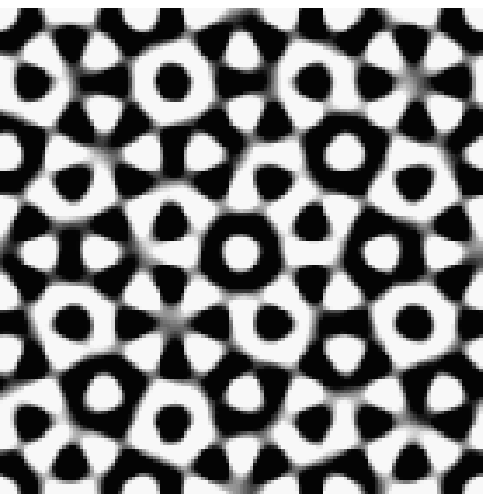}\hspace{0.5cm}
\epsfxsize=4.5cm\epsfbox{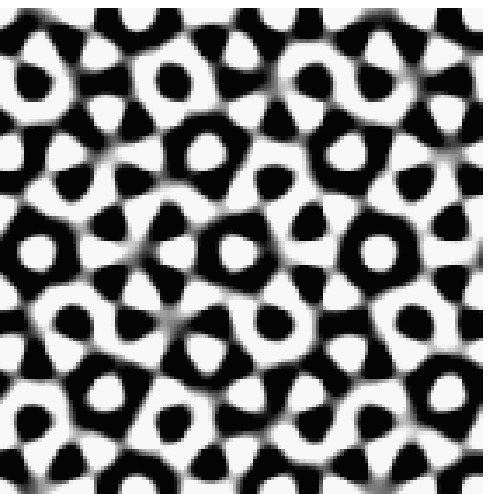}
}
\caption{Reconstruction of the octagonal quasipattern from the amplitudes
$A_i$ for the zigzag instability at the times a) t=0, b) t=1000, 
and c) t=3000. The simulations are done with $64\times 64$ Fourier modes and
the values of the coefficients: $\mu=5$, $\nu=0.7$, $\gamma=0.4$, $q=-0.05$, $L=25$, 
$k_c=10k_{min}$ and $\delta=0.1$. 
\label{fig.zz}}
\end{figure}

Assuming that the phase of each of the modes only depends on the direction
perpendicular to that mode, it is easy to see that the equations decouple.
Introducing a super-slow time, $\partial_T=\epsilon^4 \partial_{T_4}$, and taking $q$
to be small, $q=\epsilon^2 \tilde{q}$, the usual nonlinear phase equation for
the zig-zag instability is obtained:
\begin{equation}
\partial_{T_4} \phi_i = 2\tilde{q}\delta (\hat{\bf \tau}_i\cdot \nabla)^2 \phi_i
-\delta^2 (\hat{\bf \tau}_i\cdot \nabla)^4 \phi_i + 6\delta^2 
[(\hat{\bf \tau}_i\cdot \nabla)\phi_i]^2 (\hat{\bf \tau}_i\cdot \nabla)^2 \phi_i,
\end{equation}
where $\hat{\tau}_i$ is the unit vector perpendicular to $\hat{\bf n}_i$. When 
$\tilde{q}<0$ the pattern is unstable. As the coefficient in front of the cubic term
is positive, the instability is supercritical. In order to confirm this  
we have performed numerical simulations of Eqs. (\ref{eq.GLn8}) using a fourth-order
Runge-Kutta method with an integrating factor that computes the linear 
derivative terms exactly in Fourier
space. In Fig. \ref{fig.zz} the evolution of this zig-zag 
instability is shown, starting with a 
perturbation of the quasipattern in the direction perpendicular to 
$\hat{\bf n}_1$:
$A_1=Re^{iqx}(1+0.1i\cos(4\pi y/L))$, $A_j=Re^{iq\hat{\bf n}_i\cdot {\bf x}}$, $j\neq 1$. As expected, the instability saturates, resulting in a
distorted quasipattern.

\section{Decagonal Quasipatterns}

\subsection{Ginzburg-Landau Equations}

For the decagonal quasipattern we consider the expansion (cf. Fig. 
\ref{fig.star}b) $\psi = \eta \sum_{n=1}^{5}
A_n({\bf x},t) e^{i{\bf k}_n \cdot {\bf x}}+c.c.+h.o.t.$. The amplitudes 
$A_i({\bf x},t)\equiv r_i e^{i\phi_i}$  now satisfy the equations
\begin{eqnarray}
\partial_t A_i & = & \mu A_i + (\hat{\bf n}_i \cdot \tilde{\nabla})^2 A_i 
- A_i[|A_i|^2 + \nu ( |A_{i+1}|^2 + |A_{i+4}|^2) \nonumber \\
&& + \gamma ( |A_{i+2}|^2 + |A_{i+3}|^2 ) ] 
 + \alpha 
\overline{A_{i+1}}\overline{A_{i+2}}\overline{A_{i+3}}\overline{A_{i+4}}, \label{eq.GLn10}
\end{eqnarray}
where $\mu$ is the control parameter and $\nu$ and $\gamma$ measure the 
interaction between modes subtending an angle of
 $2\pi/5$ and $4\pi/5$, respectively. The
indices are repeated cyclically with period 5. Although 
the quartic term is higher order than the others, we have included 
it in Eq. (\ref{eq.GLn10}) to account for the resonant interaction among the five 
wavevectors, $\sum_{j=1}^{5} \hat{\bf k}_j =0$ (see Fig. \ref{fig.star}b).
Without 
 this term  there would be 
a spurious one-parameter class of solutions, parameterized by the global phase,
$\Phi=\sum_{j=1}^{5} \phi_j$, 
all with the same energy.  The resonance term lifts this degeneracy by rendering
the global phase (slightly) damped.

The Lyapunov functional for the dodecagonal case is given by
\begin{eqnarray}
{\cal F}&=&\int \int dxdy \sum_{i=1}^{5} \left [ -\mu |A_i|^2 
+ |(\hat{\bf n}_i \cdot \tilde{\nabla})^2 A_i|^2 + \frac{1}{2}|A_i|^4 + \nu |A_i|^2 |A_{i+1}|^2 \right .\nonumber \\
&& \left . + \gamma
|A_i|^2 |A_{i+2}|^2 \right ] -\alpha (A_1 A_2 A_3 A_4 A_5 + c.c.).
\end{eqnarray}

Writing again $A_j = a_j e^{i\phi_j}$, we now have:
\begin{eqnarray}
\partial_t a_i & = & \mu a_i  
- a_i[a_i^2 + \nu ( a_{i+1}^2 + a_{i+4}^2) 
+ \gamma ( a_{i+2}^2 + a_{i+3}^2 ) ] \nonumber \\
&& + \alpha a_{i+1}a_{i+2}a_{i+3}a_{i+4}\cos(\Phi),\\
a_i \partial_t \phi_i & = & -\alpha a_{i+1}a_{i+2}a_{i+3}a_{i+4} \sin(\Phi),
\label{eq.n10linp}
\end{eqnarray}
with $\Phi=\sum_{j=1}^{5} \phi_j$ the global phase. For a quasiperiodic solution $a_1=\cdots =a_5=R$ and Eq. (\ref{eq.n10linp})
becomes:
\begin{equation}
\partial_t \Phi = -5\alpha R^3 \sin(\Phi).
\end{equation}
There are two solutions, one stable ($\Phi=0$), the other unstable 
($\Phi=\pi$). We will consider the former one. 

Because the resonant term involving $\alpha$ is quartic, it cannot appear in
systems with the reflection symmetry $A_i \rightarrow -A_i$ which arises, for
instance, in Boussinesq convection or in Faraday waves with one-frequency forcing. 
In that case ($\alpha=0$) $\Phi$ is arbitrary to the order considered in Eq. (\ref{eq.GLn10}) and, for $\Phi \neq 0,\pi$, solutions with
pentagonal rather than decagonal symmetry can bifurcate from the trivial
state. The situation is analogous to that of the triangle solutions in the 
case of hexagonal symmetry \cite{GoSw84}. The global phase gets fixed by the 
higher-order resonance term $\overline{A}_i |A_{i+1}|^2 |A_{i+2}|^2 |A_{i+3}|^2
|A_{i+4}|^2$, which leads to $\partial_t \Phi \sim \sin(2\Phi)$, implying
$\Phi =n\pi/2$ for the regular pentagonal solution. We will not consider these solutions in the following and take $\phi_1=\cdots
=\phi_5=0$.

There are six different kinds of simple solutions (see Fig. \ref{fig.solsn5}): 
rolls, two types of rectangles, two types of one-dimensional
quasipatterns and the decagonal quasipattern. They are given by:

\begin{figure}
\centerline{
\epsfxsize=4cm\epsfbox{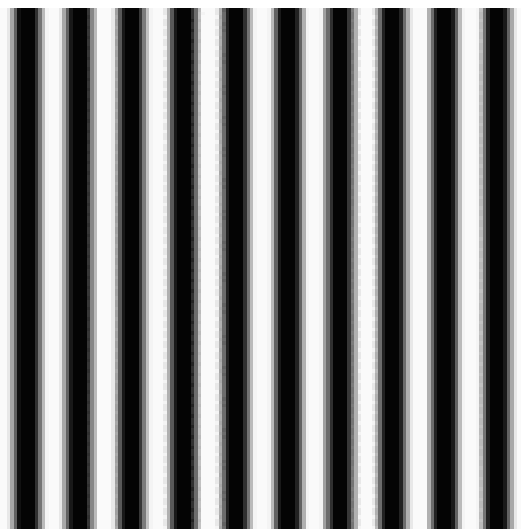}\hspace{0.5cm}
\epsfxsize=4cm\epsfbox{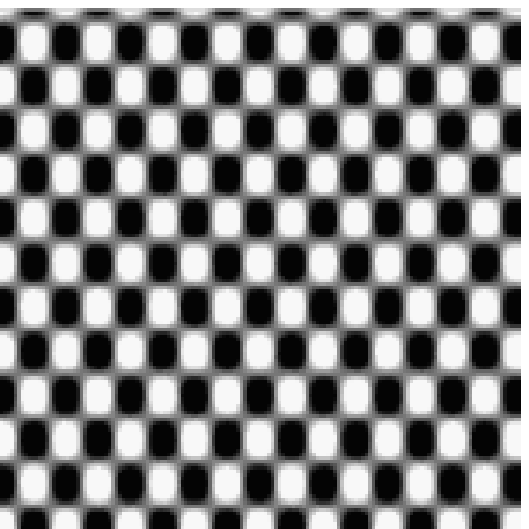}\hspace{0.5cm}
\epsfxsize=4cm\epsfbox{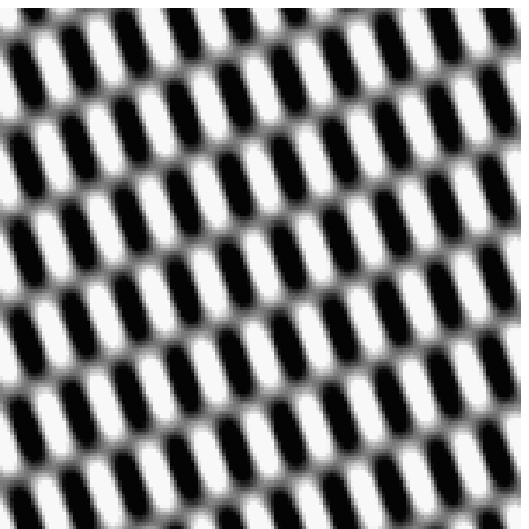}
}
\vspace{0.5cm}
\centerline{
\epsfxsize=4cm\epsfbox{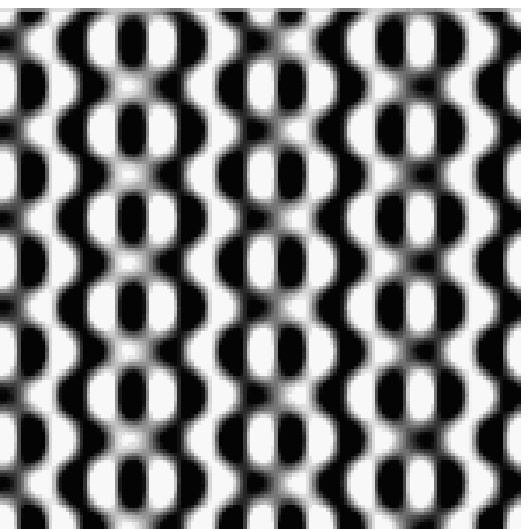}\hspace{0.5cm}
\epsfxsize=4cm\epsfbox{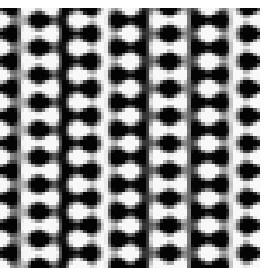}\hspace{0.5cm}
\epsfxsize=4cm\epsfbox{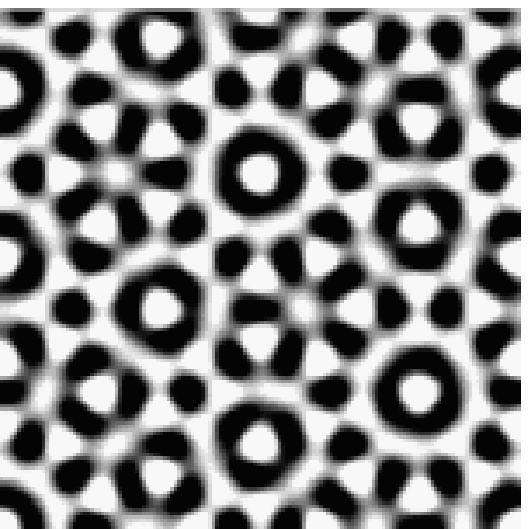}
}
\caption{Decagonal Case. 
Different types of solutions of Eqs. (\ref{eq.GLn10}). a) Rolls, b) rectangles $R_1$, c) rectangles $R_2$, d) and e) 1D quasipatterns $H_1$ and $H_2$, and f) decagonal quasipattern.\label{fig.solsn5}}
\end{figure}

\begin{enumerate}

\item Rolls: $a_2=\cdots =a_5=0$ and $a_1=\sqrt{\mu}$, $F_R=-\mu^2/2$.\\

\item Rectangles ($R_1$): $a_1=a_2=a_5=0$ and $a_3=a_4=\sqrt{\mu/(\nu+1)}$, $F_{R_1}=-\mu^2/(\nu+1)$.\\

\item Rectangles ($R_2$): $a_2=a_3=a_5=0$ and $a_1=a_4=\sqrt{\mu/(\gamma + 1)}$, $F_{R_2}=-\mu^2/(\gamma+1)$.\\

\item One-dimensional quasipatterns ($H_1$): $a_2=a_5=0$ and
\begin{equation}
a_1=\sqrt{\frac{\mu (1+\nu - 2\gamma)}{1+\nu -2\gamma^2}},\;\;
a_3=a_4=\sqrt{\frac{\mu (1-\gamma)}{1+\nu  -2\gamma^2}},
\label{eq.H1amp}
\end{equation}
\begin{equation}
F_{H_1}=-\frac{\mu^2}{2}\frac{\nu+3-4\gamma}{\nu-2\gamma^2+1}.
\end{equation}
They exist provided $(1-\gamma)(1+\nu - 2\gamma)>0$.\\ 

\item One-dimensional quasipatterns ($H_2$): $a_3=a_4=0$ and
\begin{equation}
a_1=\sqrt{\frac{\mu (1+\gamma - 2\nu)}{1+\gamma  -2\nu^2}},\;\;
a_2=a_5=\sqrt{\frac{\mu (1-\nu)}{1+\gamma -2\nu^2}},
\label{eq.H2amp}
\end{equation}
\begin{equation}
F_{H_2}=-\frac{\mu^2}{2}\frac{3+\gamma-4\nu}{1+\gamma-2\nu^2}.
\end{equation}
They exist provided $(1-\nu)(1+\gamma - 2\nu)>0$. Both, $H_1$ and $H_2$ 
are quasiperiodic in one dimension and periodic in the other.  \\

\item Quasipattern: $a_1=\cdots a_5=R$, with $R$ satisfying
\begin{equation}
0=\mu-(1+2\nu +2\gamma)R^2 + \alpha R^3. \label{eq.Qsoln10}
\end{equation}
The analytic solution is quite complicated. For $\alpha=0$ it simplifies to
\begin{equation}
R=\sqrt{\frac{\mu}{1+2\nu +2\gamma}}\;\;\;\Rightarrow\;\;\;F_Q=-\frac{5}{2}\frac{\mu^2}{1+2\nu+2\gamma}.
\end{equation}

\item In addition, when $\alpha=0$ there is another solution, with $a_5=0$ and
\begin{eqnarray}
&&a_1=a_4=\sqrt{\frac{\mu(1-\nu)}{1+\gamma+\nu+\gamma\nu-\nu^2-\gamma^2}},
\label{eq.n10.qp1}\\
&&a_2=a_3=\sqrt{\frac{\mu(1-\gamma)}{1+\gamma+\nu+\gamma\nu-\nu^2-\gamma^2}}.
\label{eq.n10.qp2}
\end{eqnarray}
It is also quasiperiodic in the two spatial dimensions and exists if 
$(1-\nu)(1-\gamma)>0$.

\end{enumerate}
 
The conditions for linear stability are:

\begin{enumerate}

\item  Rolls: $\nu>1$, $\gamma>1$.\\

\item Rectangles $R_1$: $-1<\nu<1$, $\gamma>1$.\\

\item Rectangles $R_2$: $\nu>1$, $-1<\gamma<1$.\\

\item 1D quasipattern $H_1$:
\begin{eqnarray}
&&2\gamma -1<\nu<1,\\
&&1-\nu \frac{1+\sqrt{5}}{2} - \gamma \frac{1-\sqrt{5}}{2} < 0\label{eq.ineqH1}
\end{eqnarray}

\item 1D quasipattern $H_2$:
\begin{eqnarray}
&&2\nu -1<\gamma<1,\\
&&1-\nu \frac{1-\sqrt{5}}{2} - \gamma \frac{1+\sqrt{5}}{2} < 0\label{eq.ineqH2}
\end{eqnarray}

\item The quasiperiodic solution given by Eqs. (\ref{eq.n10.qp1}), 
(\ref{eq.n10.qp2}) is always unstable.

\item Decagonal quasipattern:
\begin{eqnarray}
&&1-\nu \frac{1+\sqrt{5}}{2} - \gamma \frac{1-\sqrt{5}}{2} + R\alpha > 0, \label{ineqQn10.1}\\
&&1-\nu \frac{1-\sqrt{5}}{2} - \gamma \frac{1+\sqrt{5}}{2} + R\alpha > 0,\label{ineqQn10.2}\\
&&2(1+2\nu + 2\gamma) -3\alpha R >0. \label{eq.snQ10}
\end{eqnarray}
When (\ref{eq.snQ10}) is violated the quasipattern undergoes a saddle-node
bifurcation that generates an unstable branch with larger amplitude (see Fig. 
\ref{fig.bifn10}). For these amplitudes the quartic term is of the same order as the other 
terms in Eq. (\ref{eq.GLn10}), which is inconsistent with the amplitude expansion. 
We restrict ourselves therefore to the range $3\alpha R \ll 2(1+2\nu+2\gamma)$.

\begin{figure}
\centerline{
\epsfxsize=6cm\epsfbox{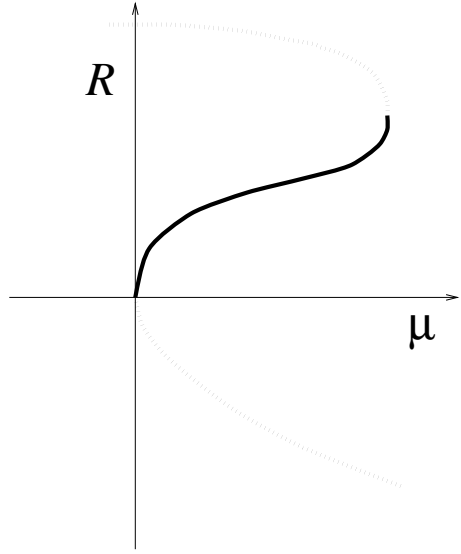}
}
\caption{Sketch of the bifurcation diagram for the decagonal quasipattern, as given by
Eq. (\ref{eq.Qsoln10}) (solid and 
dotted lines representing stable and unstable branches, respectively). The saddle-node
bifurcation occurs at amplitudes for which (\protect{\ref{eq.GLn10}}) is not valid any more.
\label{fig.bifn10}}
\end{figure} 

For $R=0$ Eqs. (\ref{ineqQn10.1},\ref{ineqQn10.2}) are complementary to the
stability conditions (\ref{eq.ineqH1}) and (\ref{eq.ineqH2}) for the one-dimensional
quasipatterns. Thus, at onset one of the one-dimensional quasipatterns
can be stable while the two-dimensional quasipattern is unstable (or vice versa). 
For $\alpha >0$, the two-dimensional quasipattern can gain stability with increasing $\mu$ while
the one-dimensional quasipattern is still stable. This bistability is expected to 
persist even if higher-order terms were  
included in the Ginzburg-Landau equations (\ref{eq.GLn10}), 
since they would contribute only terms of the
order $R^2$ to the stability conditions. For a small range of parameters 
$\nu$ and $\gamma$ the stabilization of the two-dimensional quasipattern can occur
for amplitudes for which these higher-order terms would still be 
negligible.

\end{enumerate}

\subsection{Longwave analysis}

Proceeding as in the case of the octagonal quasipattern we obtain for the
linearized perturbation equations
\begin{eqnarray}
\partial_t r_i &=& -2qR(\hat{\bf n}_i \cdot \tilde{\nabla})\phi_i + 
(\hat{\bf n}_i \cdot \tilde{\nabla})^2 r_i + R^3 \alpha (\sum_{j=1}^{5} r_j -2 r_i) \\
&&- 2 R^2 r_i - 2\nu R^2 (r_{i+1} + r_{i-1}) 
- 2\gamma R^2 (r_{i+2} + r_{i-2}),  \nonumber \\
\partial_t \phi_i & = & (\hat{\bf n}_i \cdot \tilde{\nabla})^2 \phi_i + \frac{2q}{R}(\hat{\bf n}_i \cdot \tilde{\nabla})r_i - \alpha R^3 
\sum_{j=1}^{5} \phi_j,
\end{eqnarray}
with $R$ satisfying
\begin{equation}
0=(\mu - q^2) - (1 + 2\nu + 2\gamma) R^2 + \alpha R^3.
\label{eq.Qsoln10q}
\end{equation}

For the amplitude perturbations there are three eigenvalues. One
eigenvalue, $\sigma_R = R^2 ( 3\alpha R - 2 - 4\nu - 4\gamma)$, corresponds 
to a one-dimensional eigenspace spanned by 
\begin{equation}
v_H^T=\frac{1}{\sqrt{5}}[r_1=1,r_2=1,r_3=1,r_4=1,r_5=1].
\end{equation}
The other two eigenvalues are given by $\sigma_{T2,3}= R^2 (-2 - 2
\alpha R + \nu (1 \pm \sqrt{5}) + \gamma (1 \mp \sqrt{5})$ and each 
corresponds to two two-dimensional eigenspaces. Four orthonormal vectors spanning these
spaces are
\begin{eqnarray}
&&v_{T_1}^{T}=\frac{1}{\sqrt{5+\sqrt{5}}}
\left [-(1+\sqrt{5})/2,(1+\sqrt{5})/2,-1,0,1\right ],\\ 
&&v_{T_3}^{T}=\sqrt{\frac{1}{40}}\left 
[-(1-\sqrt{5}),-(1-\sqrt{5}),-(1+\sqrt{5}),4,-(1+\sqrt{5})\right ], 
\end{eqnarray}
corresponding to $\sigma_{T2}$ and
\begin{eqnarray}
&&v_{T_2}^{T}=\frac{1}{\sqrt{5-\sqrt{5}}}\left 
[-1,(1-\sqrt{5})/2,-(1-\sqrt{5})/2,1,0\right ],\\ 
&&v_{T_4}^{T}=\sqrt{\frac{1}{40}}\left 
[-(1-\sqrt{5}),-(1+\sqrt{5}),-(1+\sqrt{5}),-(1-\sqrt{5}),4\right ], 
\end{eqnarray}
to $\sigma_{T3}$.

The global phase is a stable mode with a
one-dimensional subspace spanned by $u_{\Phi}^{T}=[\phi_1=1,\dots,\phi_5=1]/\sqrt{5}$. As with the octagonal
quasipattern,  
there are four marginal phase modes. The two corresponding to translations in
space can be written as 
\begin{eqnarray}
&&u_{\phi_x}^{T}=\sqrt{\frac{2}{5}}\left [1, \cos(2\pi/5), 
\cos(4\pi/5), \cos(6\pi/5), \cos(8\pi/5)\right ],\\
&&u_{\phi_y}^{T}=\sqrt{\frac{2}{5}}\left [0, \sin(2\pi/5),
\sin(4\pi/5), \sin(6\pi/5), \sin(8\pi/5)\right ]. 
\end{eqnarray}
The two phason modes can be written as
\begin{eqnarray}
&&u_{\varphi_1}^{T}=\sqrt{\frac{2}{5}}\left [1,\cos(6\pi/5),\cos(12\pi/5),\cos(18\pi/5),\cos(24\pi/5)\right ],\\ 
&&u_{\varphi_2}^{T} = 
\sqrt{\frac{2}{5}}\left [0,\sin(6\pi/5),\sin(12\pi/5),\sin(18\pi/5),\sin(24\pi/5)\right ]. 
\end{eqnarray}
Under rotations the phason mode $\tilde{\varphi}=(\varphi_1,\varphi_2)$ changes with 
twice the rotation angle.

The expansion of the perturbations includes now the global phase,
\begin{eqnarray}
r_i({\bf x},t) &=& \epsilon \sum_{j=1}^{4} \hat{r}^i_{j}(\epsilon^{1/2} {\bf x},
\epsilon t) v^i_{\hat{r}_j}\\
\phi_i({\bf x},t) &=& \epsilon^{1/2} [ \phi_x(\epsilon^{1/2} {\bf x},
\epsilon t) u^i_{\phi_x} + \phi_y(\epsilon^{1/2} {\bf x},
\epsilon t) u^i_{\phi_y} \nonumber \\
&& + \varphi_1(\epsilon^{1/2} {\bf x},
\epsilon t) u^i_{\varphi_1} + \varphi_2(\epsilon^{1/2} {\bf x},
\epsilon t) u^i_{\varphi_2} ]+\epsilon \Phi(\epsilon^{1/2} {\bf x},
\epsilon t) u^i_{\Phi}, 
\end{eqnarray}

and the longwave equations can be written as:
\begin{eqnarray}
\partial_t \phi_x&=&D_1\nabla^2 \phi_x + D_2 \partial_x (\nabla 
\cdot \vec{\phi}) + D_3 (\partial^2_{x^2}\varphi_1 + 2\partial^2_{xy} \varphi_2 - \partial^2_{y^2} \varphi_1),\\
\partial_t \phi_y&=&D_1\nabla^2 \phi_y + D_2 \partial_y (\nabla 
\cdot \vec{\phi}) + D_3 (\partial^2_{x^2}\varphi_1 - 2\partial^2_{xy} \varphi_1 - \partial^2_{y^2} \varphi_2),\\
\partial_t \varphi_1 &=& D_4 \nabla^2 \varphi_1 +
D_3 (\partial^2_{x^2} \phi_x - 2 \partial^2_{xy} \phi_y 
-\partial^2_{y^2} \phi_x),\\
\partial_t \varphi_2 &=& D_4 \nabla^2 \varphi_2 +
D_3 (\partial^2_{x^2} \phi_y + 2 \partial^2_{xy} \phi_x 
-\partial^2_{y^2} \phi_y).
\end{eqnarray}
The values of the coefficients are given by
\begin{eqnarray}
&&D_1 = \frac{1}{4} - \frac{q^2}{u_1},\\
&&D_2 = \frac{1}{2} - \frac{2 q^2}{v},\\
&&D_3 = D_1,\\
&&D_4= \frac{1}{2} - \frac{q^2}{u_1} - \frac{q^2}{u_2}= D_1 + \tilde{D}_4, 
\;\;\mbox{with}\;\;\tilde{D}_4 = \frac{1}{4} - \frac{q^2}{u_2},
\end{eqnarray}
with $u_1=R^2(2(R\alpha+1) -\nu ( 1+\sqrt{5}) - \gamma (1-\sqrt{5}))$, 
$u_2=R^2(2(R\alpha+1) -\nu ( 1-\sqrt{5}) - \gamma (1+\sqrt{5}))$, $v=R^2 (2 (1+2\nu + 2\gamma) -3R\alpha)$, and $R$ a solution
of Eq. (\ref{eq.Qsoln10q}).

In complex form ($\phi =
\phi_x + \rm{i} \phi_y$, $\varphi = \varphi_1 + \rm{i} \varphi_2$,
$\nabla=\partial_x + \rm{i}\partial_y$) the phase equations read
\begin{eqnarray}
&&\partial_t \phi = D_1 |\nabla|^2 \phi + \frac{1}{2}D_2 \nabla 
(\overline{\nabla} \phi + \nabla \overline{\phi}) + D_3 e^{\rm{i}\alpha} 
\overline{\nabla}^2 \overline{\varphi}, \\
&&\partial_t \varphi = D_4 |\nabla|^2 \varphi + D_3 e^{\rm{i}\alpha} 
\overline{\nabla}^2 \overline{\phi}.
\end{eqnarray}

Again, the angle $\alpha$ depends on the relative orientation between the
phonon and phason modes. For the  basis chosen, $\alpha=0$.

Considering normal modes $\phi=\phi_0 e^{i{\bf Q}\cdot {\bf x}}$, 
$\varphi=\varphi_0 e^{i{\bf Q}\cdot {\bf x}}$, the eigenvalues now become:
\begin{eqnarray}
&&\sigma_{1,2} = -\frac{1}{2}\left [ D_1 + D_4 \pm \sqrt{(D_1 - D_4)^2 + 
4D_3^2}\right ] Q^2, \\
&&\sigma_{3,4} = -\frac{1}{2}\left [ D_1 + D_2 + D_4 \pm \sqrt{(D_1 + D_2 - 
D_4)^2 + 4D_3^2}\right ] Q^2. 
\end{eqnarray}

\begin{figure}
\centerline{
\epsfxsize=7cm\epsfbox{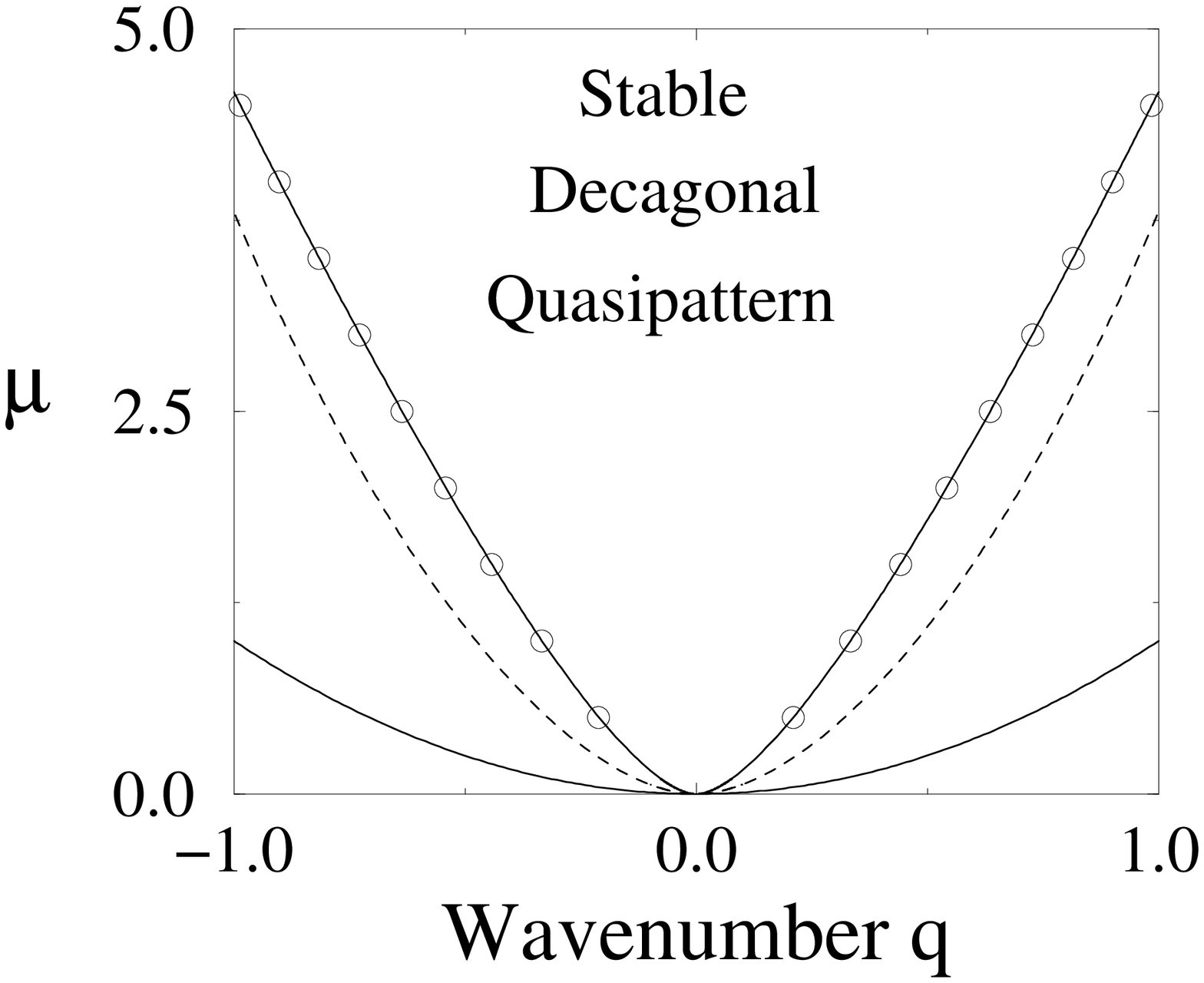}\hspace{0.5cm}
\epsfxsize=7cm\epsfbox{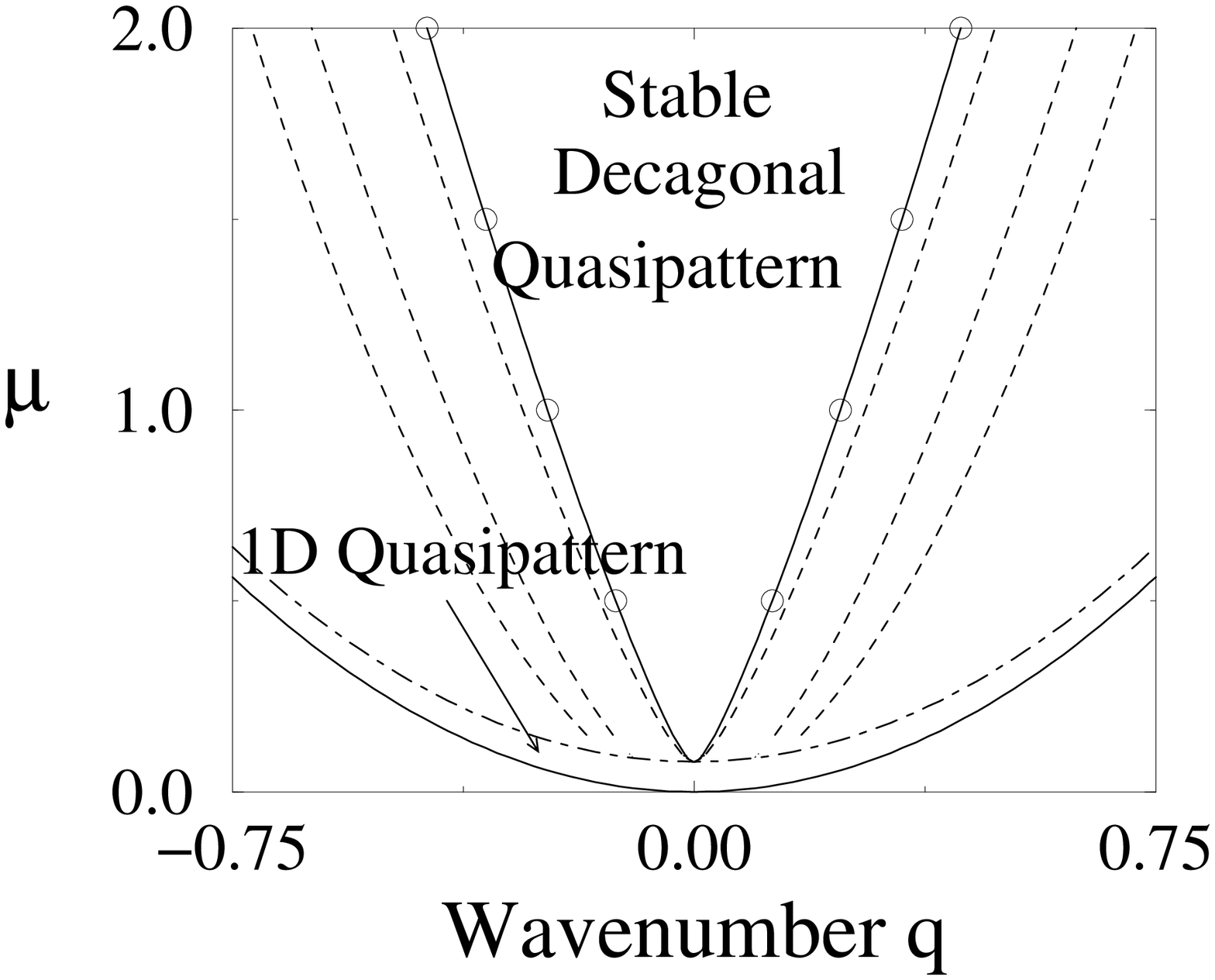}
}
\caption{Stability diagrams for $\alpha=1$ and a) $\nu=\gamma=0.7$, 
b) $\nu=0.5$, $\gamma=0.9$. 
The dash-dotted line represents the line at which the
2-d quasipattern becomes stable. Below this line, 
the 1-d quasipattern $H_2$ is the only stable solution.
\label{fig.stabn10}}
\end{figure}

Typical stability diagrams are shown in Fig. \ref{fig.stabn10}. When $\nu=\gamma$ 
the eigenvalues become degenerate ($D_4=2D_1$ and three eigenvalues go through
zero at the curve $u_1=u_2=4q^2$.) In Fig.\ref{fig.stabn10}b
the one-dimensional quasipattern persists amplitude-stable beyond the dashed-dotted lines
(cf.(\ref{eq.ineqH2})). We have not investigated its stability with respect to 
sideband perturbations. 

\section{Dodecagonal Quasipatterns}

\subsection{Ginzburg-Landau Equations}

Dodecagonal quasipatterns can be thought of as a combination of two rotated
hexagon patterns. The amplitudes $A_i({\bf x},t)$ corresponding to the 
modes indicated in Fig. \ref{fig.star}c satisfy the equations:
\begin{eqnarray}
\partial_t A_i & = & \mu A_i + (\hat{\bf n}_i \cdot \tilde{\nabla})^2 A_i 
 + \alpha\overline{A_{i+2}}\overline{A_{i+4}}
- A_i[|A_i|^2 + \nu ( |A_{i+2}|^2 + |A_{i+4}|^2) \nonumber \\
&& + \gamma ( |A_{i-1}|^2 + |A_{i-3}|^2 ) +2\beta |A_{i+1}|^2 ] 
,\;\;\; i=1,3,5 \label{eq.GLn12.1},\\
\partial_t A_i & = & \mu A_i + (\hat{\bf n}_i \cdot \tilde{\nabla})^2 A_i 
 + \alpha\overline{A_{i+2}}\overline{A_{i+4}}
- A_i[|A_i|^2 + \nu ( |A_{i+2}|^2 + |A_{i+4}|^2) \nonumber \\
&& + \gamma ( |A_{i+1}|^2 + |A_{i+3}|^2 ) +2\beta |A_{i-1}|^2 ] 
,\;\;\; i=2,4,6. \label{eq.GLn12.2}
\end{eqnarray}
Here the indices are cyclic with period 6, $\mu$ is the 
control parameter and $\nu$, $\gamma$ and $\beta$ measure 
the interaction between modes subtending angles of 
$2\pi/3$, $\pi/6$ and $\pi/2$, respectively. The two hexagonal sub-lattices 
imply two resonance conditions ${\bf k}_1 + {\bf k}_3 + {\bf k}_5 =0$ and
${\bf k}_2 + {\bf k}_4 + {\bf k}_6 =0$. Correspondingly, there are two global
phases: $\Phi_1=\phi_1 + \phi_3 + \phi_5$ and $\Phi_2=\phi_2 + \phi_4 + 
\phi_6$.

The Lyapunov functional can be written as:
\begin{eqnarray}
{\cal F}&=&\int \int dxdy \sum_{i=1}^{6} \left [ -\mu |A_i|^2 
+ |(\hat{\bf n}_i \cdot \tilde{\nabla})^2 A_i|^2 + \frac{1}{2}|A_i|^4 + \nu |A_i|^2 |A_{i+2}|^2   \right. \nonumber \\
&&\left. + \frac{\gamma}{2}
|A_i|^2 |A_{i+3}|^2 + \frac{\gamma+2\beta}{2}|A_i|^2 |A_{i+1}|^2 + 
\frac{(-1)^{i+1}(\gamma-2\beta)}{2}|A_i|^2 |A_{i-1}|^2) 
\right ] \nonumber \\ 
&&-\alpha (A_1 A_3 A_5 +  A_2 A_4 A_6 + c.c.).
\end{eqnarray}

In terms of the magnitudes $a_i$ and phases $\phi_i$ 
Eqs. (\ref{eq.GLn12.1},\ref{eq.GLn12.2}) can be written as
\begin{eqnarray}
\partial_t a_i & = & \mu a_i  + \alpha a_{i+2}a_{i+4}\cos(\Phi)
 \nonumber \\
&& - a_i[a_i^2 + \nu ( a_{i+2}^2 + a_{i+4}^2) 
+ \gamma ( a_{i\pm 1}^2 + a_{i\pm 3}^2 ) + 2\beta |a_{i\pm 1}|^2 ],\\
a_i \partial_t \phi_i & = & -\alpha a_{i+2}a_{i+4} \sin(\Phi), \label{eq.n12linp}
\end{eqnarray}
where $\Phi=\Phi_1$ or $\Phi=\Phi_2$ for $i$ odd and $i$ even, respectively.
For hexagonal or dodecagonal patterns in which the non-zero amplitudes are
fixed, $a_i=R$, Eq. (\ref{eq.n12linp}) becomes 
\begin{equation}
\partial_t \Phi_{1,2} = -3\alpha R \sin(\Phi_{1,2}).\\
\end{equation}
Among the four solutions $\Phi_{1,2}=0$ and $\Phi_{1,2}=\pi$ only the one with $\Phi_1=\Phi_2=0$ is stable. In the following we 
assume the amplitudes $A_j$ to be 
real and take $\phi_1=\cdots
=\phi_6=0$. Now the simple solutions include rolls, rectangles, squares, hexagons,
mixed modes, one-dimensional quasipatterns and the dodecagonal quasipattern
(see Fig. \ref{fig.solsn12}). They are given by:

\begin{figure}
\centerline{
\epsfxsize=4cm\epsfbox{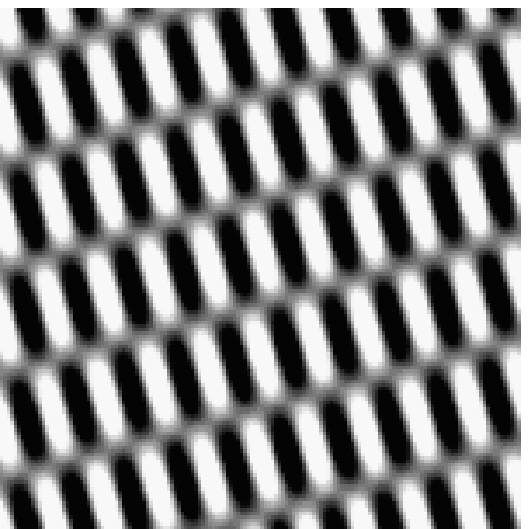}\hspace{0.5cm}
\epsfxsize=4cm\epsfbox{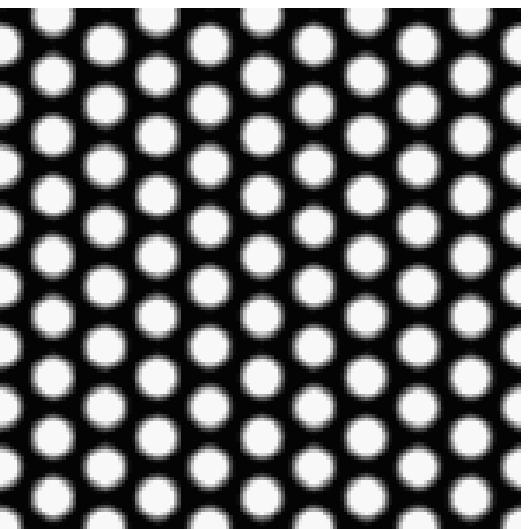}\hspace{0.5cm}
\epsfxsize=4cm\epsfbox{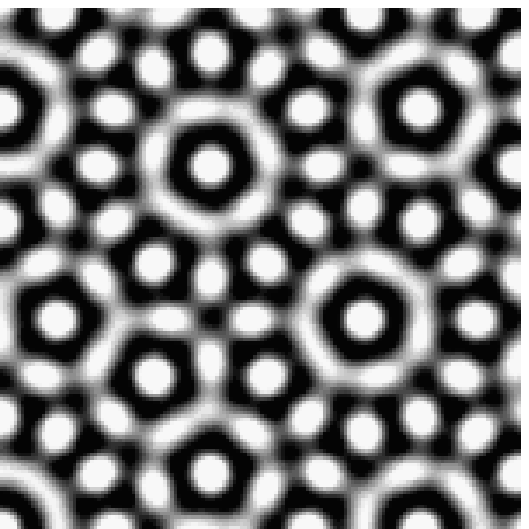}
}
\caption{Dodecagonal Case. Several of the solutions of Eqs. (\ref{eq.GLn12.1}), 
(\ref{eq.GLn12.2}). a) Rectangles, b) hexagons, and d) dodecagonal quasipattern.\label{fig.solsn12}}
\end{figure}

\begin{enumerate}

\item Rolls: $a_2=\cdots =a_6=0$ and $a_1=\sqrt{\mu}$, $F_R=-\mu^2/2$.\\

\item Rectangles ($R$): $a_2=\cdots=a_5=0$ and $a_1=a_6=\sqrt{\mu/(\gamma+1)}$, $F_{R_1}=-\mu^2/(\gamma+1)$.\\

\item Squares ($S$): $a_3=\cdots=a_6=0$ and $a_1=a_2=\sqrt{\mu/(2\beta + 1)}$, $F_{R_2}=-\mu^2/(2\beta+1)$.\\

\item Hexagons ($H$): $a_2=a_4=a_6=0$ and
\begin{equation}
a_1=a_3=a_5=\frac{\alpha\pm\sqrt{\alpha^2 +4\mu(1+2\nu)}}{2(1+2\nu)},
\end{equation}
\begin{equation}
F_{H}=-\frac{3}{2}R^2[2\mu-R^2 (1+2\nu)]-\alpha R^3.
\end{equation}

\item Mixed mode, $a_2=a_4=a_6=0$, $a_1=a_3 \neq a_5$. 
It is always unstable.\\

\item One-dimensional quasipattern, $a_2=a_4=0$, $a_{1,3,5,6} \neq 0$.\\

\item Two-dimensional Quasipattern, $a_1=\cdots a_6=R$, with $R$ given by,
\begin{equation}
R=\frac{\alpha \pm \sqrt{\alpha^2 + 
4\mu (1+2\nu+2\gamma+2\beta)}}{2(1+2\nu+2\gamma+2\beta)}, \label{eq.Qsoln12}
\end{equation}
\begin{equation}
F_{Q}=-3R^2[2\mu-R^2 (1+2\nu+2\gamma+2\beta)]-2\alpha R^3.
\end{equation}

\end{enumerate}

The relative stability of the former solutions is as follows:

\begin{enumerate}

\item Rolls are always unstable at onset. Provided $\nu>1$, $\beta>1/2$, 
and $\gamma>1$, they become stable at a larger value of $\mu$, given by
\begin{equation}
\mu=\frac{\alpha^2}{(\nu-1)^2}.
\end{equation}

\item Rectangles are unstable at onset and become stable at  
\begin{equation}
\mu=\frac{\alpha^2 (1+\gamma)}{(\nu-1)(\nu+2\beta-\gamma-1)},
\end{equation}
if $\nu>1$ and $\nu+2\beta > \gamma+1 >0$.

\item Squares are also unstable at onset. If $\nu + \gamma > 1+2\beta >0$ they
become stable at
\begin{equation}
\mu=\frac{\alpha^2 (1+2\beta)}{(1+2\beta-\nu-\gamma)^2}.
\end{equation}

\item Hexagons appear in a saddle-node bifurcation at
\begin{equation}
\mu=-\frac{\alpha^2}{4(1+2\nu)},
\end{equation}
but when $2(\gamma + \beta) < -(1+2\nu)$ they are always unstable with respect 
to the dodecagonal quasipattern. If 
$2(\gamma + \beta) > -(1+2\nu)$ they are stable at the saddle-node
but can become unstable as a 
result of a secondary bifurcation. In particular, if $\nu > 1$ they can become 
unstable to rolls at
\begin{equation}
\mu=\frac{\alpha^2 (2+\nu)}{(\nu -1)^2},
\end{equation}
and, when $1+2\nu > 2(\gamma + \beta)$, to the dodecagonal quasipattern for
\begin{equation}
\mu > \frac{2\alpha^2 (\gamma + \beta)}{(1+2\nu -2\gamma-2\beta)^2}
\end{equation}.

 
\item Dodecagonal quasipatterns appear in a saddle-node bifurcation at
\begin{equation}
\mu=-\frac{1}{4}\frac{\alpha^2}{(1+2\nu+2\gamma+2\beta)^2}.
\end{equation}
When $2|\gamma + \beta| < 1+2\nu$, they are unstable at onset with respect to 
hexagons, but become stable through a secondary bifurcation, at 
\begin{equation}
\mu=\frac{\alpha^2}{4}\frac{6\gamma+6\beta-2\nu-1}{(1+2\nu-2\gamma-2\beta)^2}.
\label{eq.stablQ}
\end{equation}
Therefore, there is hysteresis between hexagons and the dodecagonal quasipattern.
A similar situation is found in the case of superlattices \cite{SiPr98}.

Besides, if $1+\gamma < \nu +2\beta$ or $1+2\beta < \gamma + \nu$ the 
quasipattern becomes again unstable to rectangles or squares, respectively. The 
values of the control parameter for these transitions are
\begin{equation}
\mu_{rec}=\frac{\alpha^2(3\gamma+\nu+2)}{(1+\gamma-\nu-2\beta)^2}.
\end{equation}
and
\begin{equation}
\mu_{sq}=\frac{\alpha^2(2+\gamma+\nu+4\beta)}{(1+2\beta-\gamma-\nu)^2}.
\end{equation}

\end{enumerate}

\subsection{Longwave analysis}

The longwave analysis proceeds analogous to that of the previous cases. The 
linearized perturbation equations are given by:
\begin{eqnarray}
\partial_t r_i &=& -2qR(\hat{\bf n}_i \cdot \nabla)\phi_i + 
(\hat{\bf n}_i \cdot 
\nabla)^2 r_i + R \alpha (r_{i+2} + r_{i+4} - r_{i}) \\
&&- 2 R^2 r_i - 2\nu R^2 (r_{i+2} + r_{i+4}) 
- 2\gamma R^2 (r_{i\pm 1} + r_{i\pm 3})- 4\beta R^2 r_{i\mp 1},  \nonumber \\
\partial_t \phi_i & = & (\hat{\bf n}_i \cdot \nabla)^2 \phi_i + \frac{2q}{R}(\hat{\bf n}_i \cdot \nabla)r_i - \alpha R
 (\phi_i + \phi_{i+2} + \phi_{i+4}),
\end{eqnarray}
with
\begin{equation}
R=\frac{\alpha+ \sqrt{\alpha^2 + 
4(\mu-q^2)(1+2\nu+2\gamma+2\beta)}}{2(1+2\nu+2\gamma+2\beta)}.
\end{equation}

For the amplitude perturbations there are four eigenvalues. Two correspond 
to one dimensional eigenspaces, $\sigma_{T_1} = -R [ 2R(1 + 2\nu + 
2\gamma + 2\beta)-\alpha]$, with eigenvector $v_H^T=[1,1,1,1,1,1]/\sqrt{6}$ and 
$\sigma_{T_2} = -R [ 2R(1 + 2\nu - 2\gamma - 2\beta)-\alpha]$,
with $v_{T_1}=[-1,1,-1,1,-1,1]/\sqrt{6}$. The other four eigenvalues 
$\sigma_{T_{3,4}}= -2R [R(1+\gamma -2\beta -\nu)+\alpha]$ and 
$\sigma_{T_{5,6}}= -2R [R(1+2\beta -\gamma -\nu) + \alpha]$ are associated 
with two two-dimensional eigenspaces. Four orthonormal vectors spanning these
spaces are:
\begin{equation} 
v_{T_3}^{T}=[-1/2,1/2,1/2,-1/2,0,0], \;\;v_{T_4}^{T}=[-1/2,1/2,0,0,1/2,-1/2],
\end{equation}
corresponding to $\sigma_{T_{3,4}}$ and
\begin{equation}
v_{T_5}^{T}=[1/2,1/2,0,0,-1/2,-1/2], \;\;v_{T_6}^{T}=[0,0,1/2,1/2,-1/2,-1/2]
\end{equation} 
to $\sigma_{T_{4,5}}$.

The global phases are again stable modes with 
one-dimensional subspaces spanned by $u_{\Phi_1}^{T}=[1,0,1,0,1,0]/\sqrt{3}$ 
and  $u_{\Phi_2}^{T}=[0,1,0,1,0,1]/\sqrt{3}$.
There are four marginal modes. The two phase modes 
corresponding to translations in space can be written as 
\begin{eqnarray}
&&u_{\phi_x}^{T}=[1, 0, -1/2, -\sqrt{3}/2, -1/2,\sqrt{3}/2]/\sqrt{3},\\ &&u_{\phi_y}^{T}=[0, 1,\sqrt{3}/2,-1/2,-\sqrt{3}/2,-1/2]/\sqrt{3}. 
\end{eqnarray}
An orthonormal base for the
phasons, which corresponds to a relative translation of the two hexagonal
lattices, is given by:
\begin{eqnarray}
&&u_{\varphi_1}^{T}=[1,0,-1/2,\sqrt{3}/2,-1/2,-\sqrt{3}/2]/\sqrt{3},\\    &&u_{\varphi_2}^{T} = [0,1,-\sqrt{3}/2,-1/2,\sqrt{3}/2,-1/2]/\sqrt{3}. 
\end{eqnarray}
As in the case of the octagonal quasipattern, 
under rotations the phason field $\tilde{\varphi}=(\varphi_1,\varphi_2)$ 
changes with five times the rotation angle.

An expansion very similar to that in the decagonal case leads to the
longwave equations
\begin{eqnarray}
\partial_t \vec{\phi}&=&D_1\nabla^2 \vec{\phi} + (D_2 - D_1) \nabla (\nabla 
\cdot \vec{\phi}), \\
\partial_t \varphi_1 &=& D_3 \nabla^2 \varphi_1 +
(D_4 - D_3) \partial_y (\partial_y \varphi_1 + \partial_x \varphi_2),\\
\partial_t \varphi_2 &=& D_3 \nabla^2 \varphi_2 +
(D_4 - D_3) \partial_x (\partial_x \varphi_1 + \partial_y \varphi_2),
\end{eqnarray}
with $\vec{\phi}=(\phi_x,\phi_y)$. Note that to this order the
equation for the phonon modes of the dodecagonal pattern
is the same as that for an
isotropic medium. The values of the coefficients are:
\begin{eqnarray}
&&D_1 = \frac{1}{4} - \frac{q^2}{u_1},\\
&&D_2 = \frac{3}{4} - \frac{2q^2}{v_1} - \frac{q^2}{u_1},\\
&&D_3= \frac{1}{4} - \frac{q^2}{u_2},\\
&&D_4 = \frac{3}{4} - \frac{2q^2}{v_2} - \frac{q^2}{u_2},
\end{eqnarray}
with $v_1=-R [\alpha-2R(2\beta+2\nu+2\gamma+1)]$, 
$v_2=-R[\alpha+2R(2\beta +2\gamma - 2\nu -1)]$, $u_1=2R[\alpha+R(1+\gamma-
2\beta-\nu)]$, $u_2=2R[\alpha + R(1+2\beta -\gamma-\nu)]$ and $R$ given
by Eq. (\ref{eq.Qsoln12}). 

In complex form the longwave equations become:
\begin{eqnarray}
&&\partial_t \phi = \frac{1}{2}(D_1 + D_2) |\nabla|^2 \phi + \frac{1}{2}(D_2 - D_1) \nabla 
(\overline{\nabla} \phi + \nabla \overline{\phi}),  \\
&&\partial_t \varphi = \frac{1}{2} (D_3 + D_4)|\nabla|^2 \varphi + 
\frac{1}{2}(D_3 - D_4) \overline{\nabla}^2 \overline{\varphi}.
\end{eqnarray}

\begin{figure}
\centerline{
\epsfxsize=7.5cm\epsfbox{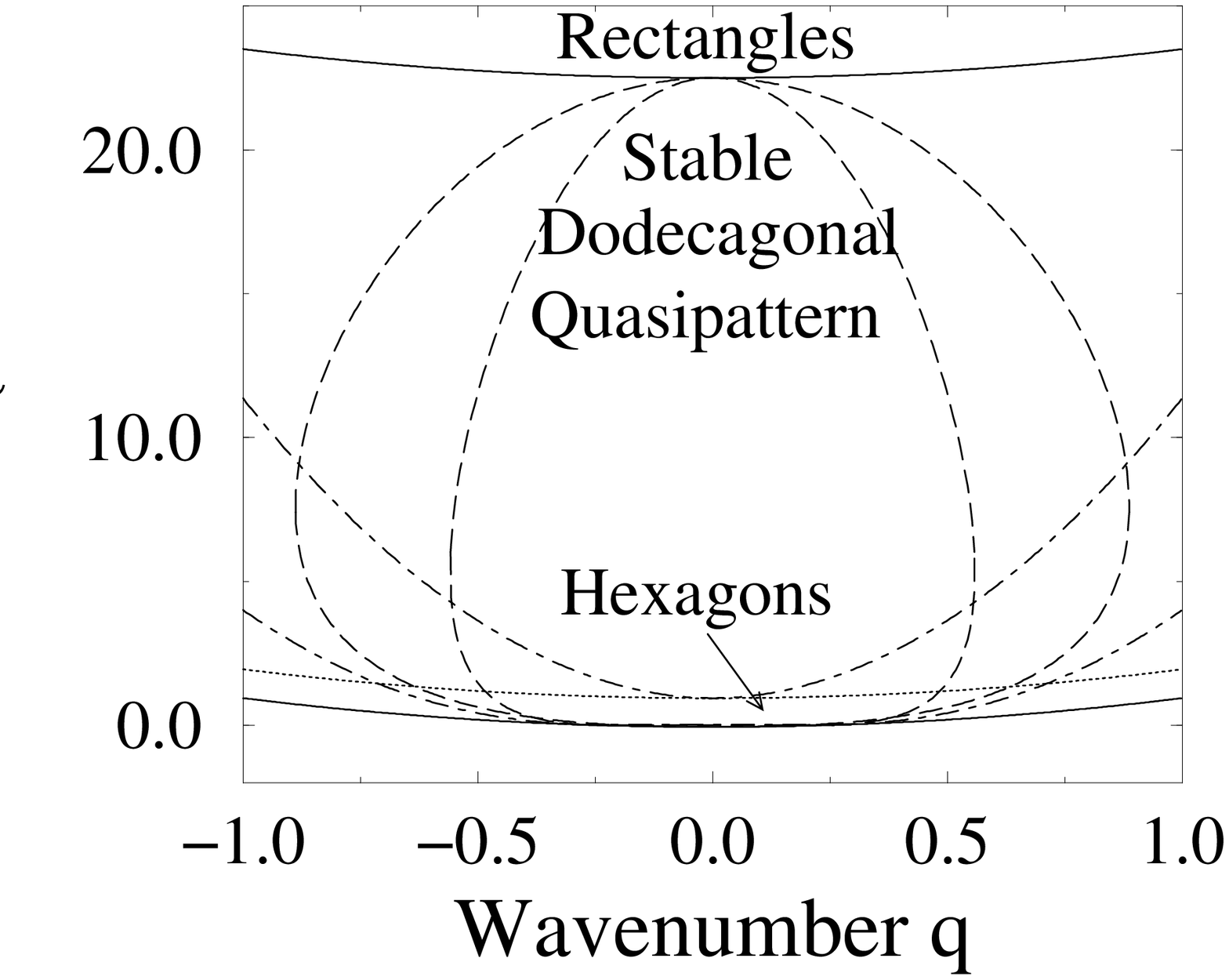}\hspace{0.2cm}
\epsfxsize=7.5cm\epsfbox{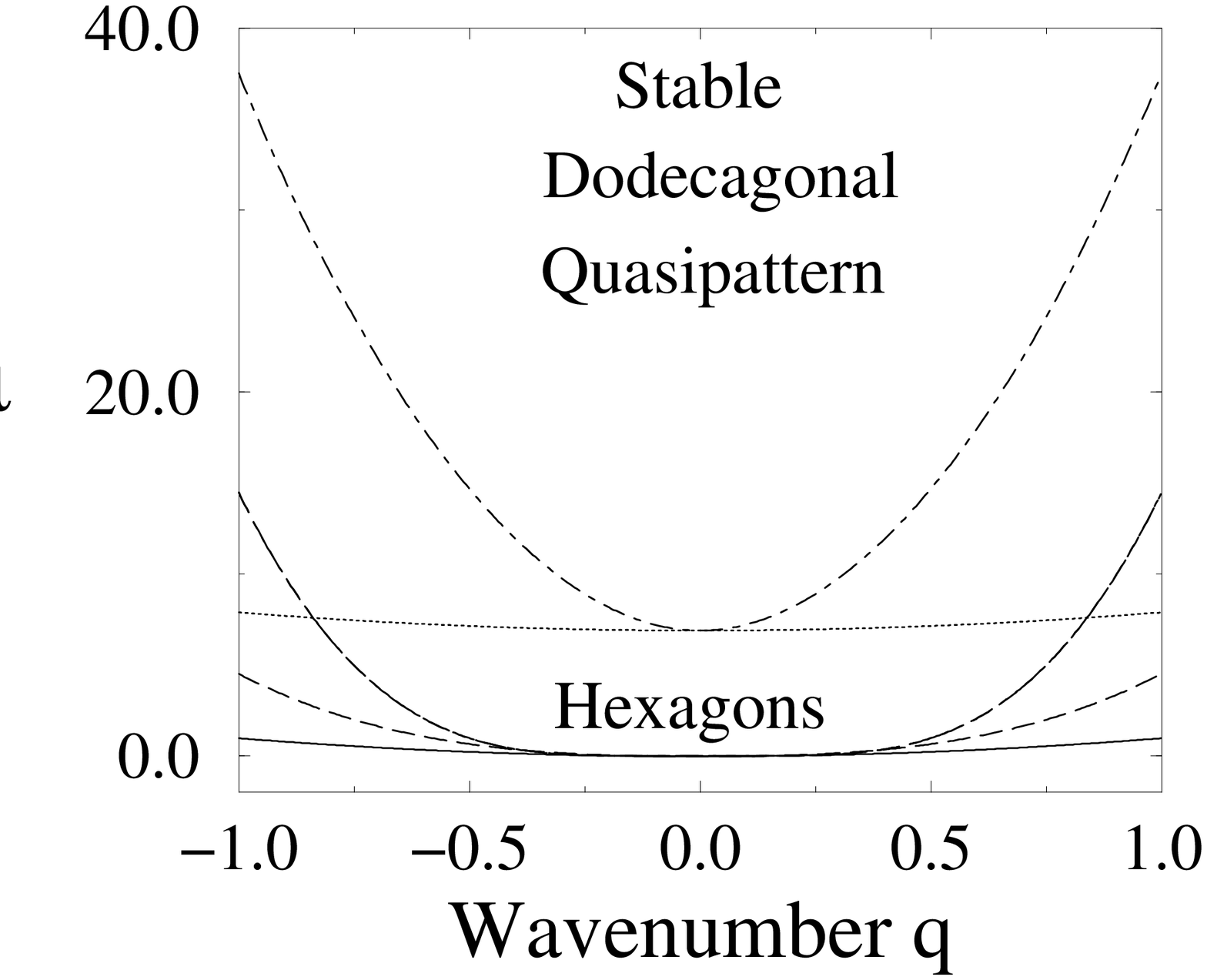}\hspace{0.2cm}
}
\vspace{1cm}
\centerline{
\epsfxsize=7.5cm\epsfbox{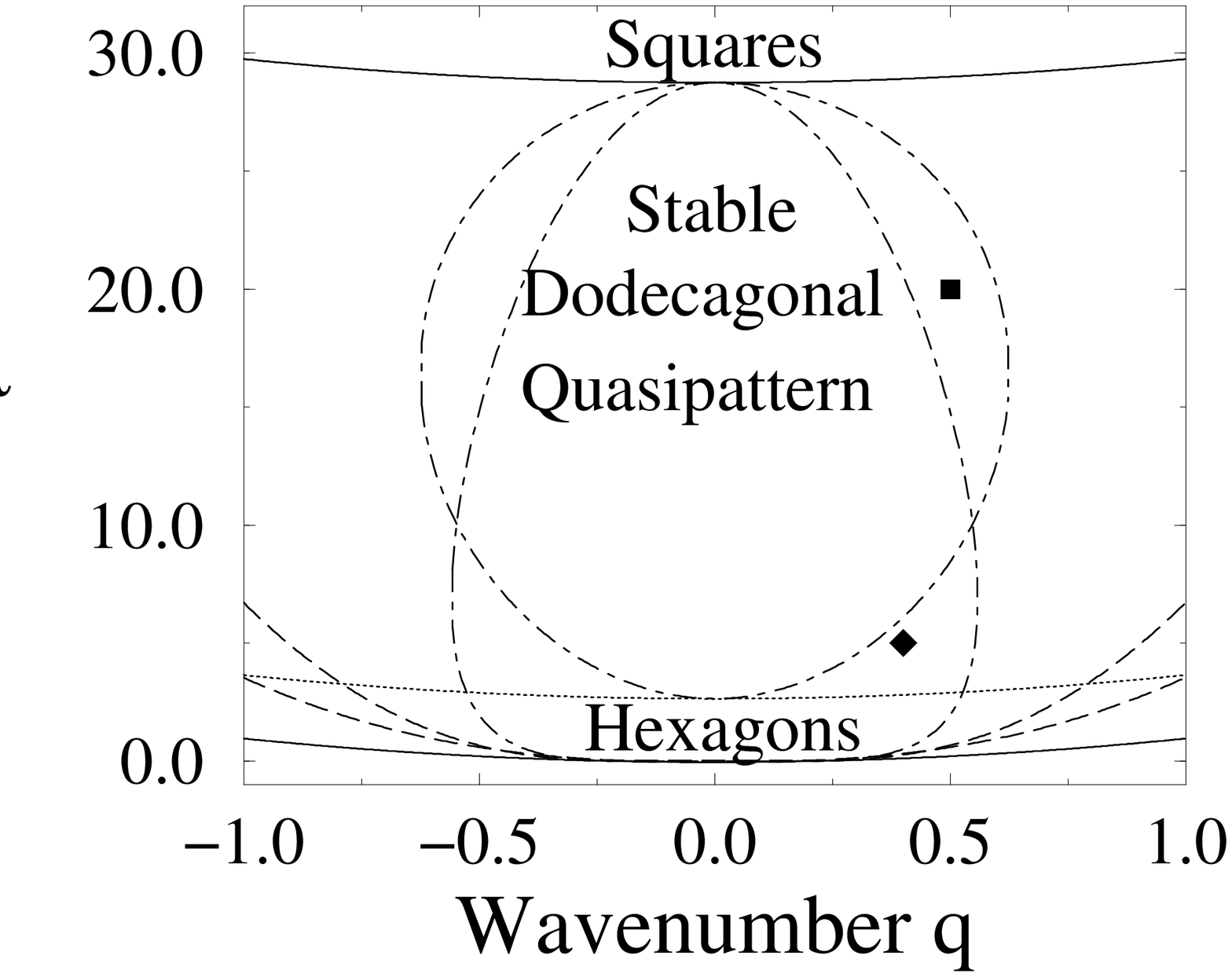}
}
\caption{Stability diagrams for $\alpha=1$ and a) $\nu=0.7$, $\gamma=0.3$ and
$\beta=0.5$, b) $\nu=0.9$, $\gamma=0.8$ and $\beta=0.4$, and c) $\nu=0.9$, $\gamma=0.9$ and $\beta=0.2$. The lower solid line corresponds to the 
saddle-node bifurcation of the quasipattern. In all the figures the 
quasipattern is unstable at onset and becomes stable above the dotted
line (cf. Eq. (\ref{eq.stablQ})). In a) and c) there is a further transition
to rectangles and squares, respectively, above the upper solid line. The 
stable region of the quasipattern is limited by longwave instabilities 
corresponding to the phonon (long dashed lines) and phason modes (dashed-dotted
lines). The diamond and square in c) correspond to the simulations in Figs. 
\ref{fig.instphD3} and \ref{fig.instphD4}, respectively.
\label{fig.stabn12}}
\end{figure}

One feature that distinguishes the dodecagonal quasipattern from the octagonal and the decagonal quasipattern is the fact that its longwave dynamics
decouple into pure phonon and pure phason modes. 
Thus, the eigenvalues are simply given by
\begin{equation}
\sigma_{1} = - D_1 Q^2, \;\;\sigma_{2} = - D_2 Q^2,\;\;
\sigma_{3} = - D_3 Q^2, \;\;\sigma_{4} = - D_4 Q^2, 
\end{equation}
and one obtains separate phase and phason instabilities when $D_{1,2}$ or $D_{3,4}$ 
change sign, respectively. 
Typical stability diagrams are shown in Fig. \ref{fig.stabn12}. The stability limits
due to phonon and phason modes are indicated by dashed and 
dashed-dotted lines, 
respectively. Over wide ranges of parameters the stable wavenumber band is limited
by phason modes.

For the phonon-type modes, the eigenvectors correspond to the usual 
irrotational and divergence free modes (equivalent to transversal and 
longitudinal waves in an elastic medium.) They satisfy:
\begin{equation}
\nabla \cdot \vec{\phi}_l=0,\;\;\; \nabla \times \vec{\phi}_t =0.
\end{equation}

For the phason modes, on the other hand, the eigenvectors are:
\begin{equation}
\tilde{\varphi}^{\sigma_3}=\left [
\begin{array}{c}
-Q_x\\
Q_y
\end{array}
\right ]e^{i{\bf Q}\cdot {\bf x}},\;\;\;
\tilde{\varphi}^{\sigma_4}=\left [
\begin{array}{c}
Q_y\\
Q_x
\end{array}
\right ]e^{i{\bf Q}\cdot {\bf x}}.
\end{equation}
These modes satisfy the equations:
\begin{equation}
\partial_y \varphi^{\sigma_3}_{1} + \partial_x \varphi^{\sigma_3}_{2} = 0,\;\;\;
\partial_x \varphi^{\sigma_4}_{1} - \partial_y \varphi^{\sigma_4}_{2}=0.
\end{equation}

\begin{figure}
\centerline{
\epsfxsize=6cm\epsfbox{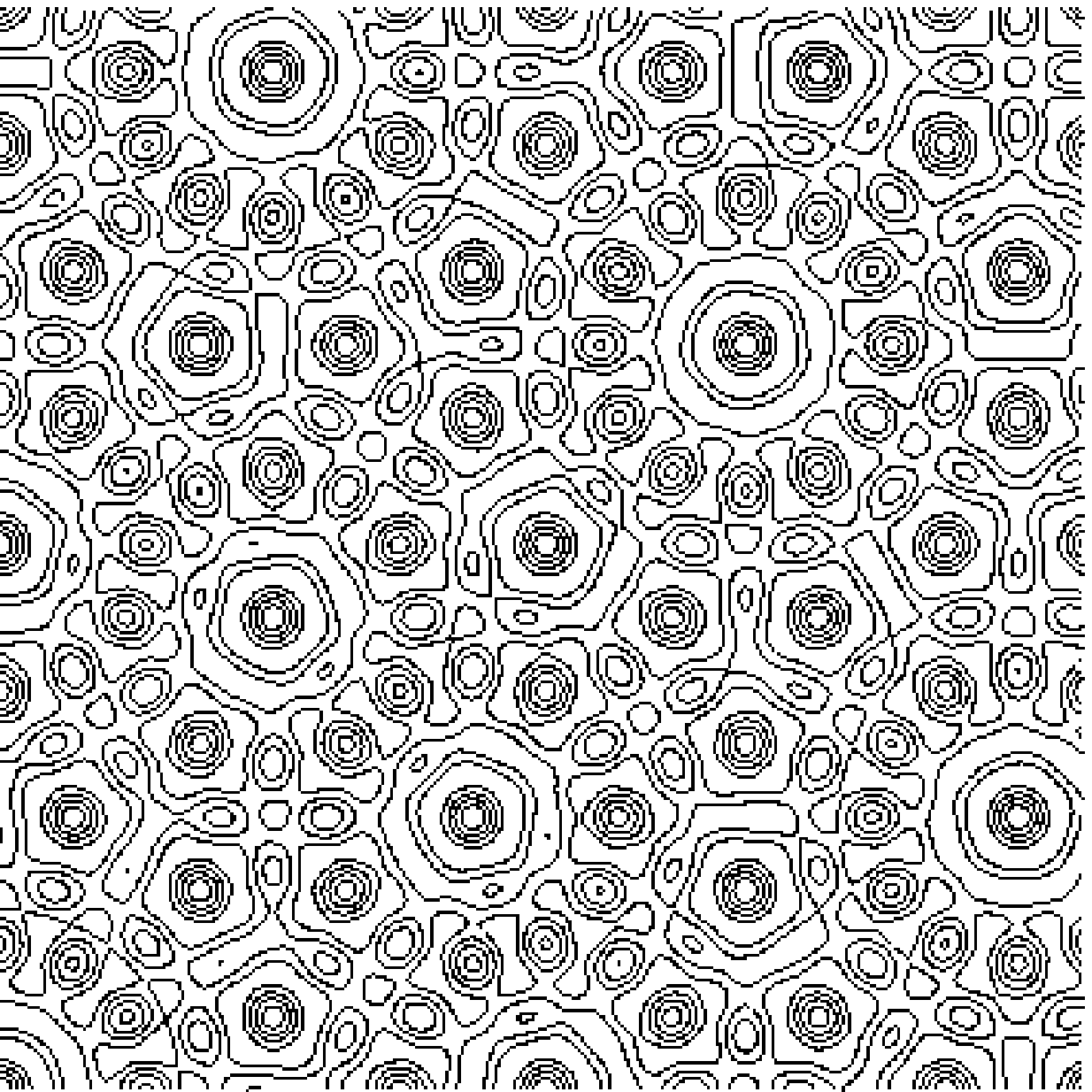}\hspace{0.5cm}
\epsfxsize=6cm\epsfbox{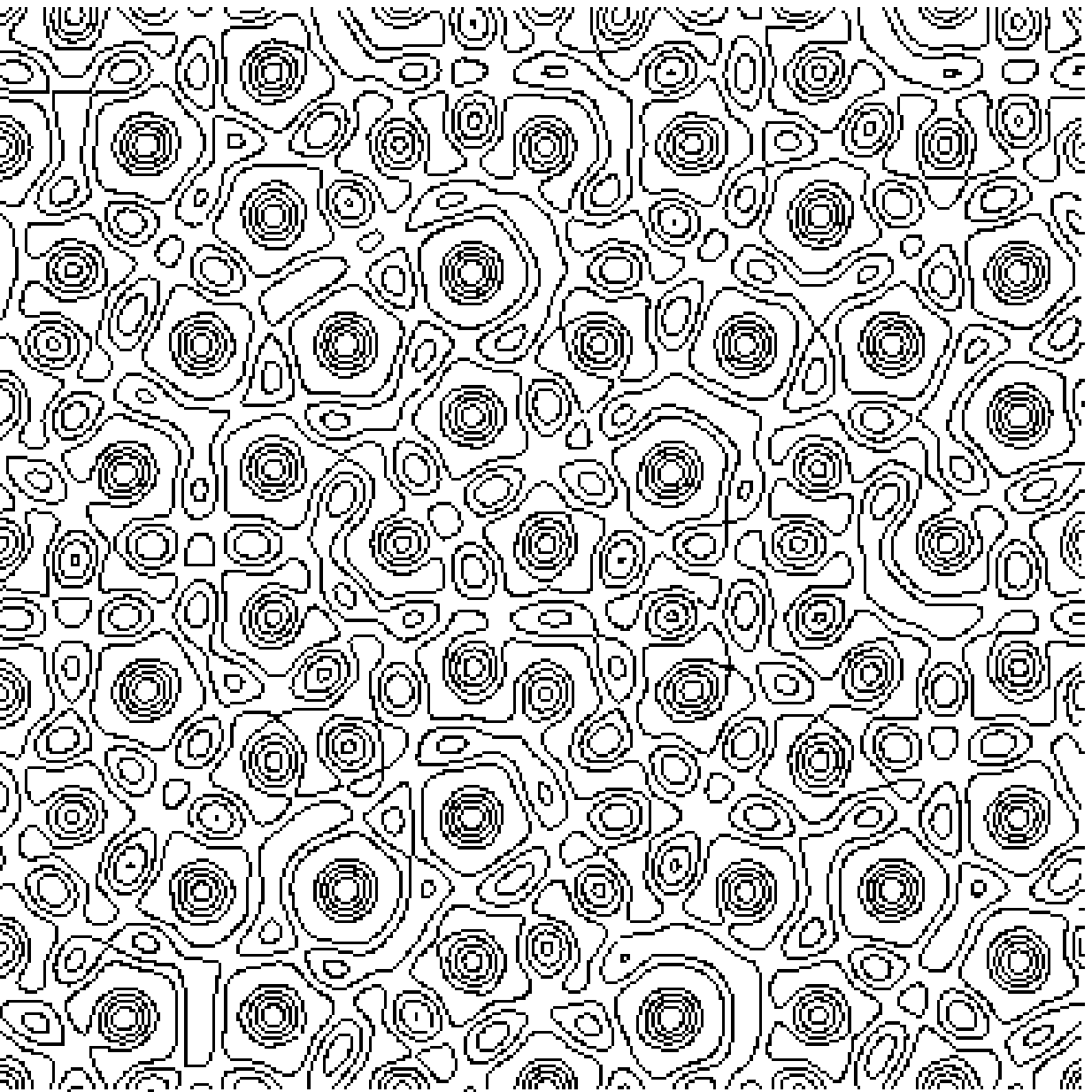}
}
\vspace{0.5cm}
\centerline{
\epsfxsize=6cm\epsfbox{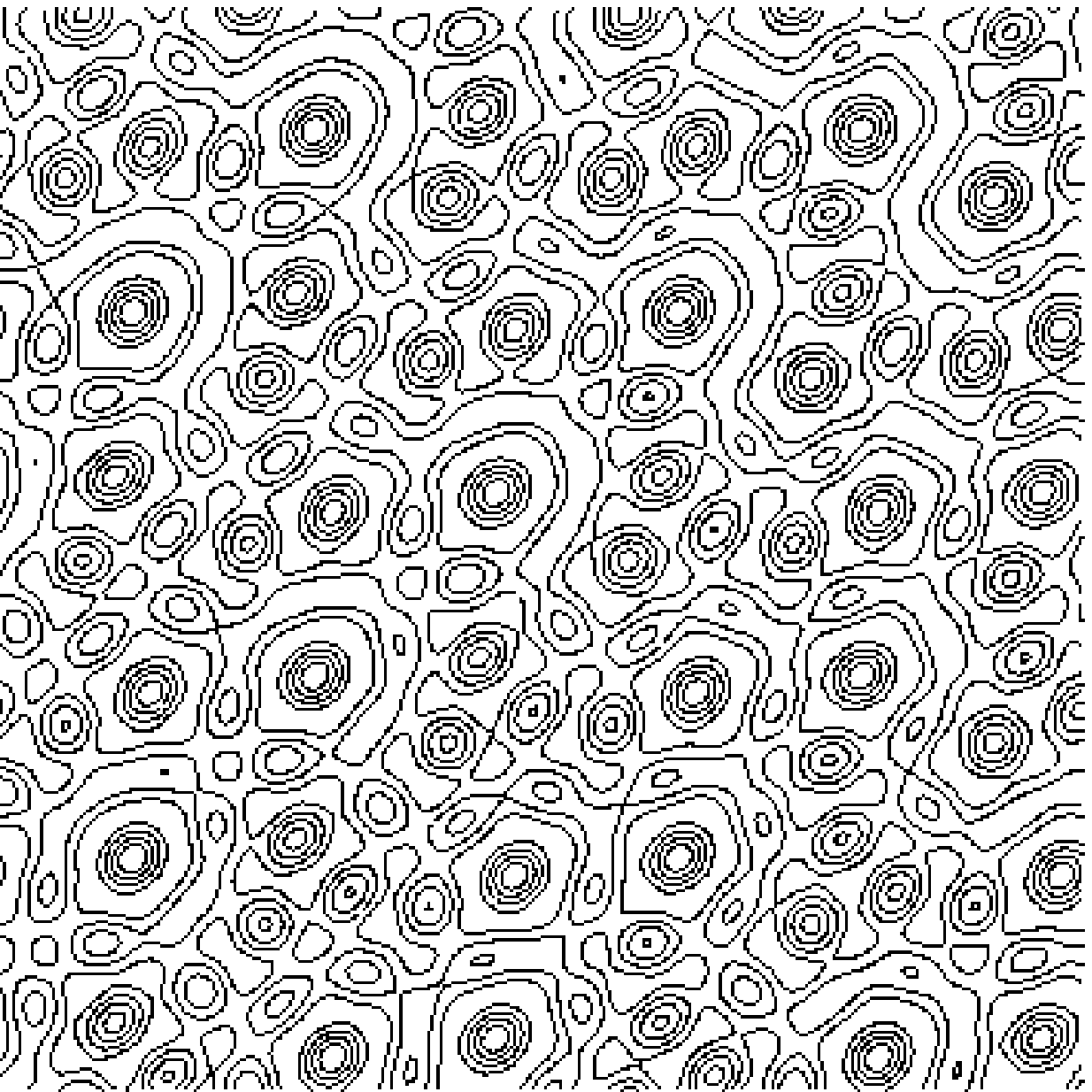}}
\caption{Instability corresponding to $\sigma_3>0$ (diamond in Fig. 
\ref{fig.stabn12}c) for a) t=0, b) t=380, and c) t=450. 
($\mu=5$, $\nu=0.9$, $\gamma=0.9$, $\beta=0.2$, $q=0.4$, $L=50$, $k_c=16k_{min}$).
\label{fig.instphD3}}
\end{figure} 

\begin{figure}
\centerline{
\epsfxsize=6cm\epsfbox{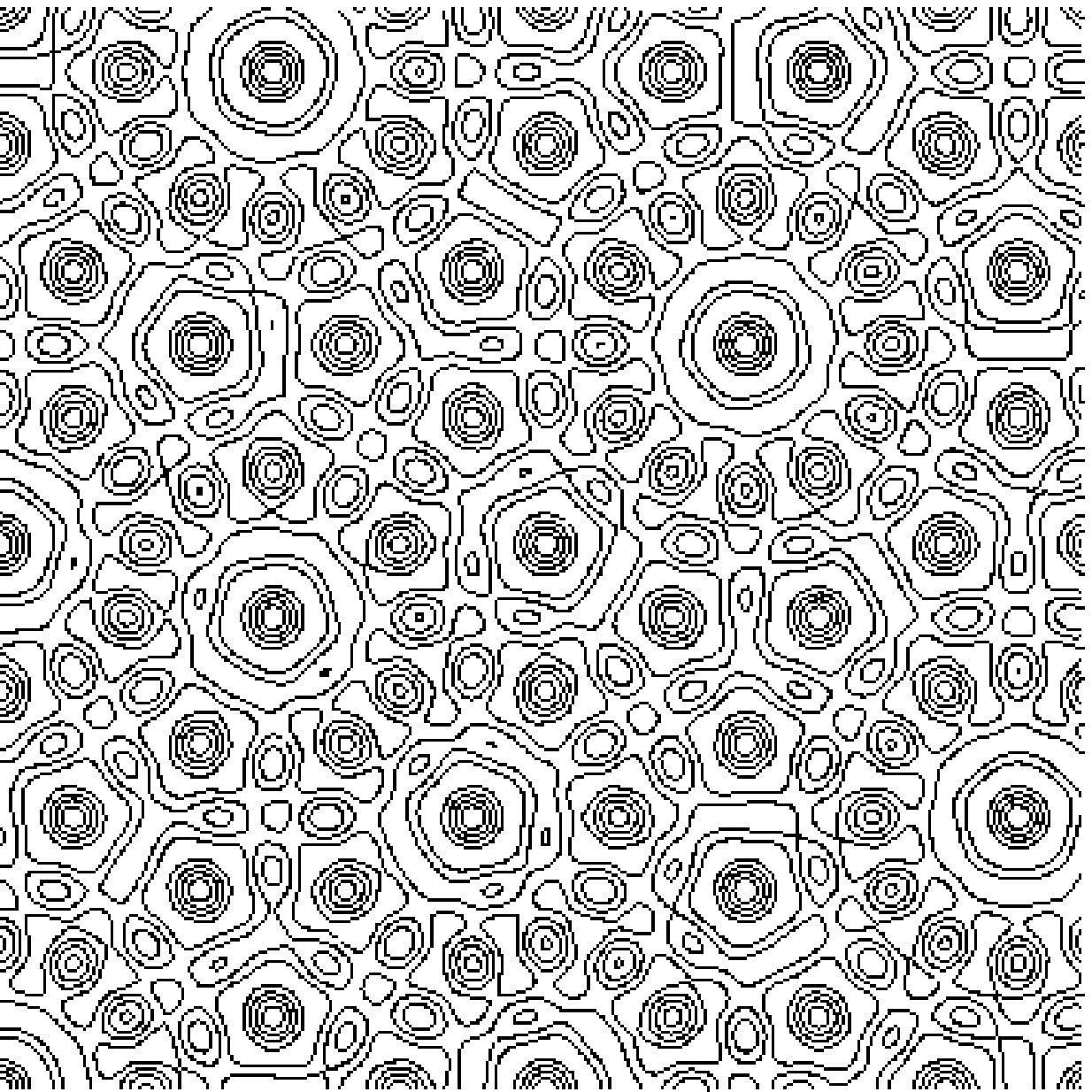}\hspace{0.5cm}
\epsfxsize=6cm\epsfbox{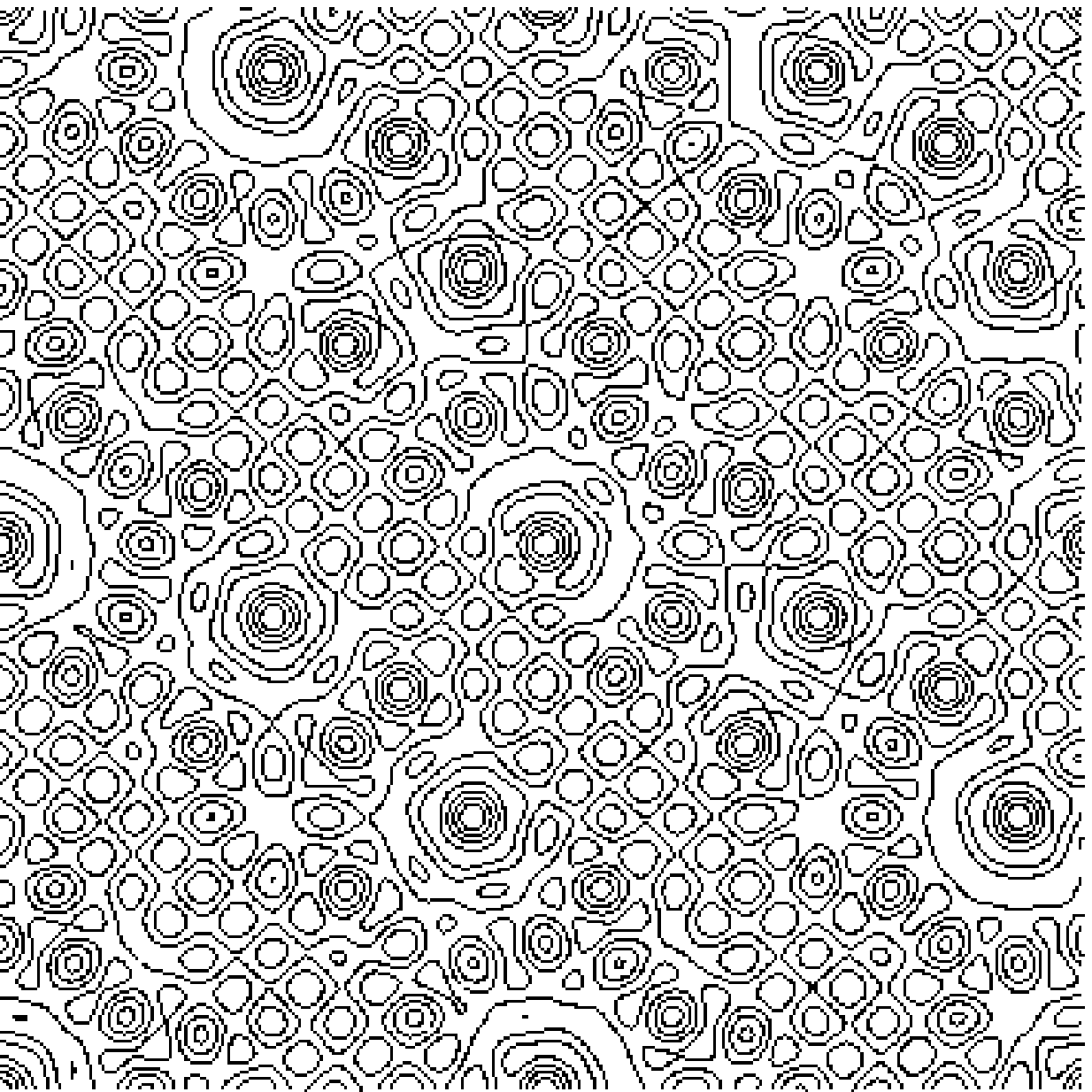}\hspace{0.5cm}
}
\vspace{0.5cm}
\centerline{
\epsfxsize=6cm\epsfbox{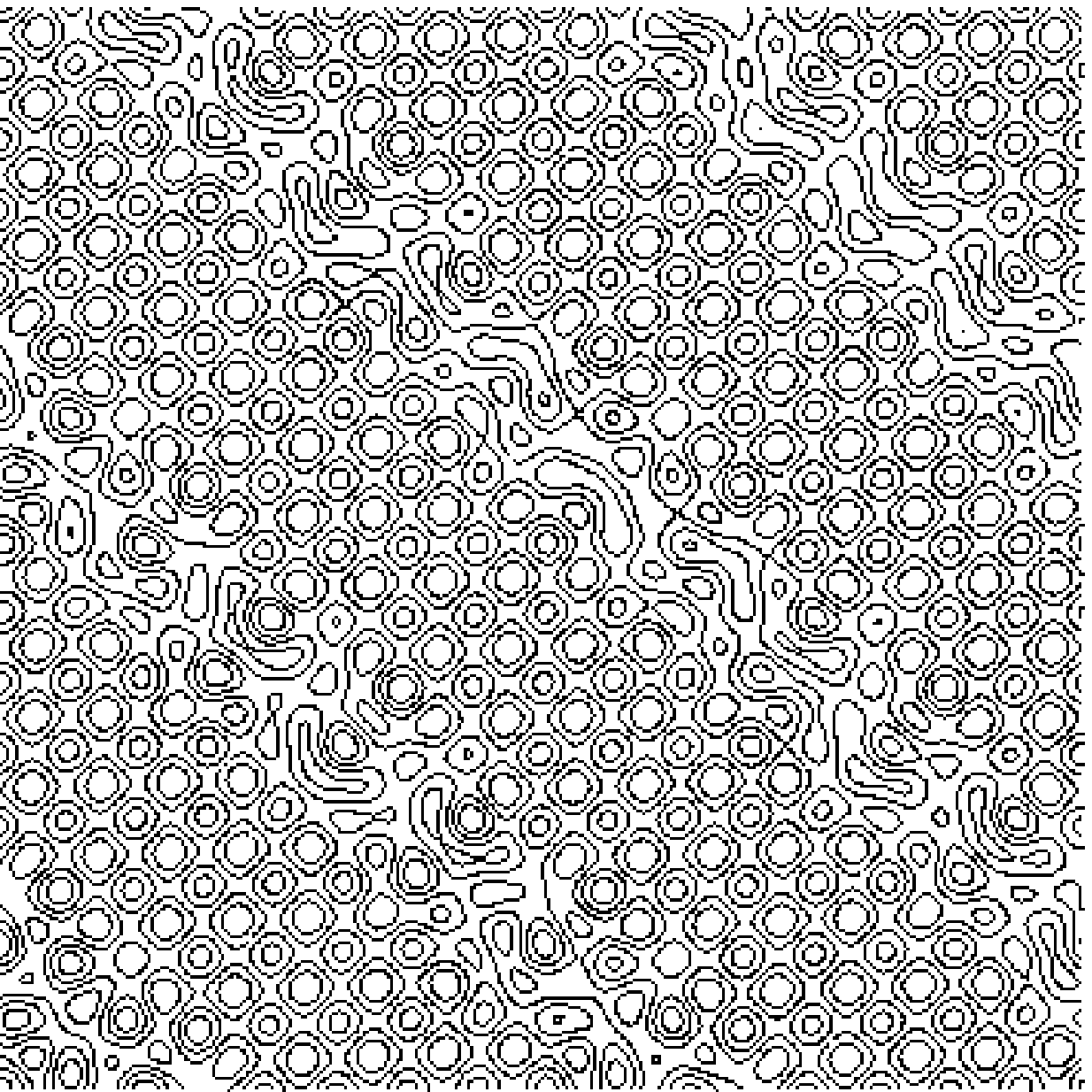}\hspace{0.5cm}
\epsfxsize=6cm\epsfbox{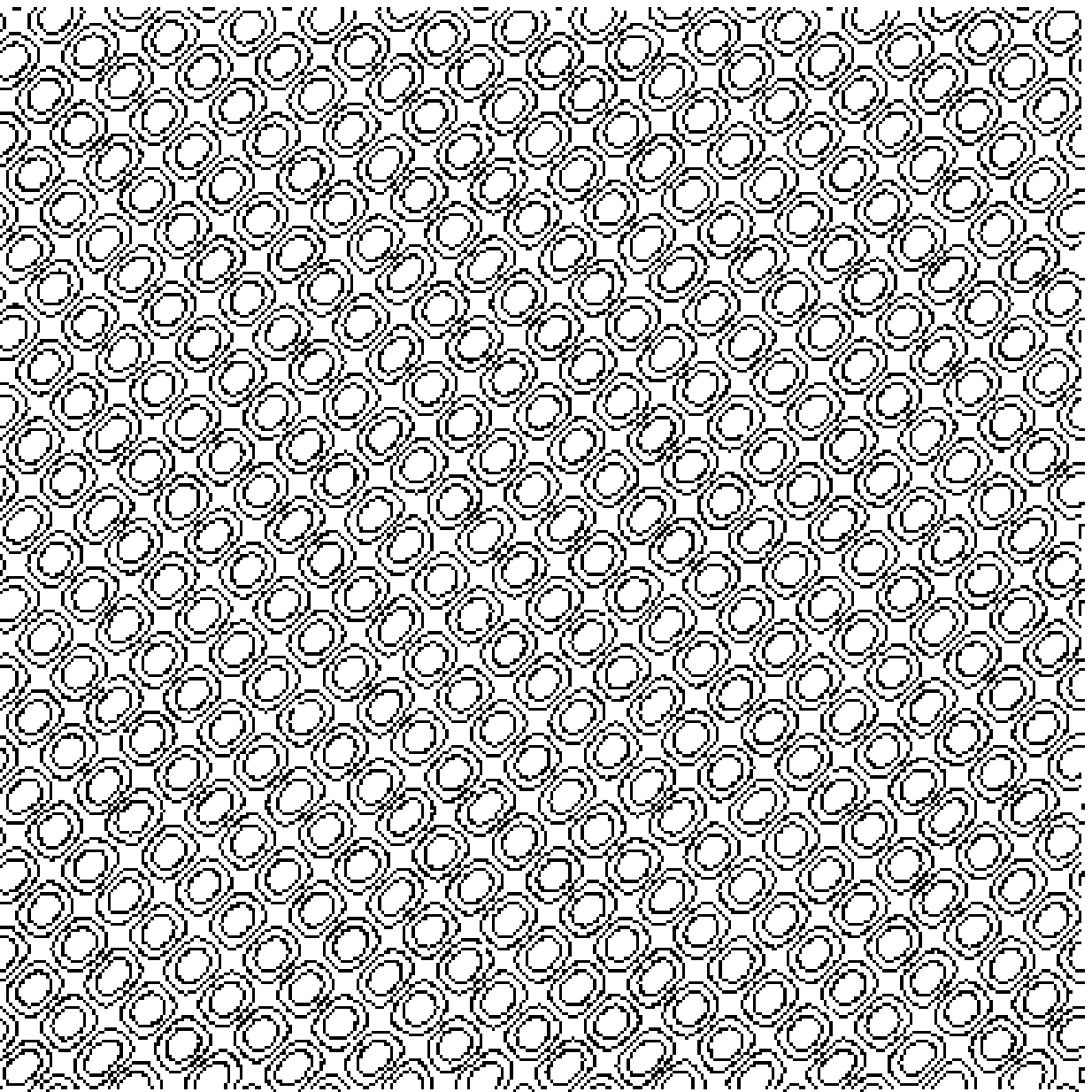}\hspace{0.5cm}
}
\caption{Instability corresponding to $\sigma_4>0$ (square in Fig. \ref{fig.stabn12}c)
for a) t=0, b) t=440, c) t=445 and d) t=600. 
($\mu=20$, $\nu=0.9$, $\gamma=0.9$, $\beta=0.2$, $q=0.5$, $L=25$, $k_c=16k_{min}$).
\label{fig.instphD4}}
\end{figure}

In order to study the behavior arising from these instabilities we have 
simulated numerically Eqs. (\ref{eq.GLn12.1}), (\ref{eq.GLn12.2}). We start 
with a perfect dodecagonal quasipattern and add a perturbation in the form of
$\tilde{\varphi}^{\sigma_3}$ (Fig. \ref{fig.instphD3}a) or 
$\tilde{\varphi}^{\sigma_4}$ (Fig. \ref{fig.instphD4}a), with 
${\bf Q}=(4\pi/L,4\pi/L)$. Since the perturbation is along the diagonal,
the evolution of the system will be quasi one-dimensional. In 
Fig. \ref{fig.instphD3} we show the evolution of the instability 
corresponding to $\sigma_3 >0$. The perturbation
grows until it creates line defects 
at time t=380 (Fig. \ref{fig.instphD3}b) along which four of the amplitudes
vanish and change their wavenumber,
demonstrating that the instability is subcritical. In Fig. \ref{fig.instphD3}c
the final state is shown, after the line defects have gone. It corresponds to
a slightly distorted quasipattern.

The instability corresponding to $\sigma_4 >0$ is also subcritical, generating 
line defects (Fig. \ref{fig.instphD4}b). But for the values of the parameters
in the simulation the square pattern minimizes the Lyapunov functional and it
nucleates around the line defects. The patches of squares grow 
(Fig. \ref{fig.instphD4}c) until they occupy the whole cell, resulting in a
slightly distorted square pattern (Fig. \ref{fig.instphD4}d).

\section{Defects}

The properties of defects in the quasipatterns differ noticeably in the 
three cases investigated in this paper. Due to the absence of resonance terms in
(\ref{eq.GLn8}) (also to higher orders) the octagonal quasipattern can be considered 
as made up of four sub-lattices corresponding to the four basic wavevectors. 
Each of the sub-lattices has its own defects. Thus, taking into account the two possible 
`charges' of the defects there are $2\times4$ different defects.
 They interact strongly with 
defects of the same sub-lattice {\it via}  the phase. The interaction  
with the defects in the other sub-lattices is only through the variations in the magnitude and
is much weaker. Overall, we expect the 
dynamics to be quite similar to that of defects in square patterns, which are made of two 
rather than four sublattices. 
Topologically, two defects of different sub-lattices could bind to 
form a vectorial defect (cf. \cite{Pi94}). However, in our simulations we have not observed 
such kind of defects in the octagonal quasipattern. 
This seems to be consistent with simulations of the Vector Complex
Ginzburg-Landau Equation, where vectorial defects were never observed in the
potential limit \cite{HeHo00}. 

For the dodecagonal quasipatterns the Ginzburg-Landau equations 
(\ref{eq.GLn12.1},\ref{eq.GLn12.2})
have (quadratic) resonance terms coupling the three modes in each of the two hexagonal
sub-lattices. At the defects therefore two amplitudes in the same sub-lattice must vanish
leading to penta-hepta defects very similar to those in the usual hexagonal patterns.
However, while in the core of penta-hepta defects of hexagons the pattern corresponds to the 
roll pattern, it corresponds in the dodecagonal case to the one-dimensional quasipattern. 
As in the hexagonal case, each penta-hepta defect carries two independent charges 
corresponding to the two vanishing amplitudes. This leads to a total of 
$(2\times3)\times(2\times2)\times2/2=12$ different defects.
While in the octagonal case collisions between defects can only annihilate them or
leave them unchanged, collisions between penta-hepta defects  
can change their type. 
An example of such a collision is given by the process
\bea
(+,+,0;0,0,0) \ \  + \ \  (-,0,+;0,0,0) \qquad \rightarrow \qquad (0,+,+;0,0,0). 
\label{e:coll}
\eea
Here $+/-$ in the $i^{th}$ entry stands for a defect with positive/negative 
charge in the mode
$A_i$ while $0$ indicates that the corresponding amplitude has no defect.
The semi-colon separates the two hexagonal sub-lattices.  
While there are a number of different such type-changing collisions,
a penta-hepta defect in one sub-lattice can never
change into one in the other sub-lattice, since there is no resonance
term in (\ref{eq.GLn12.1},\ref{eq.GLn12.2}) involving modes of 
both hexagonal sub-lattices. In other words, there is 
no penta-hepta defect that has one vanishing amplitude in one sub-lattice and 
one vanishing amplitude in the other sub-lattice.

Penta-hepta defects within the same
sub-lattice interact with each other strongly through the phase. 
The strength of the interaction (attraction $vs.$ repulsion)
is expected to be related to the sum $N$ of the products of the charges of the 
individual defects \cite{Ts96},
\bea
N=\sum_{j=1}^{n} \delta_j^1 \delta_j^2. \label{e:Ndef}
\eea
Here $\delta^{1,2}_j$ is the topological charge of the first and 
second defect, respectively, in the mode $A_j$. 
For hexagons ($n=3$) $N$ can only take on the values -2, -1, 1, and 2, since for any 
defect pair there is always a mode that vanishes in both of them. Thus, within each 
sublattice all penta-hepta defects interact strongly. Penta-hepta 
defects of different sub-lattice will interact only weakly (through the magnitude). 
Considering that collisions do not change defect type across the 
sublattices, one may expect the ordering dynamics that leads from disordered patterns to 
ordered quasipatterns to occur in the two sub-lattices essentially independently. 

\begin{figure}

\centerline{
\epsfxsize=3.5cm\epsfysize=2cm\epsfbox{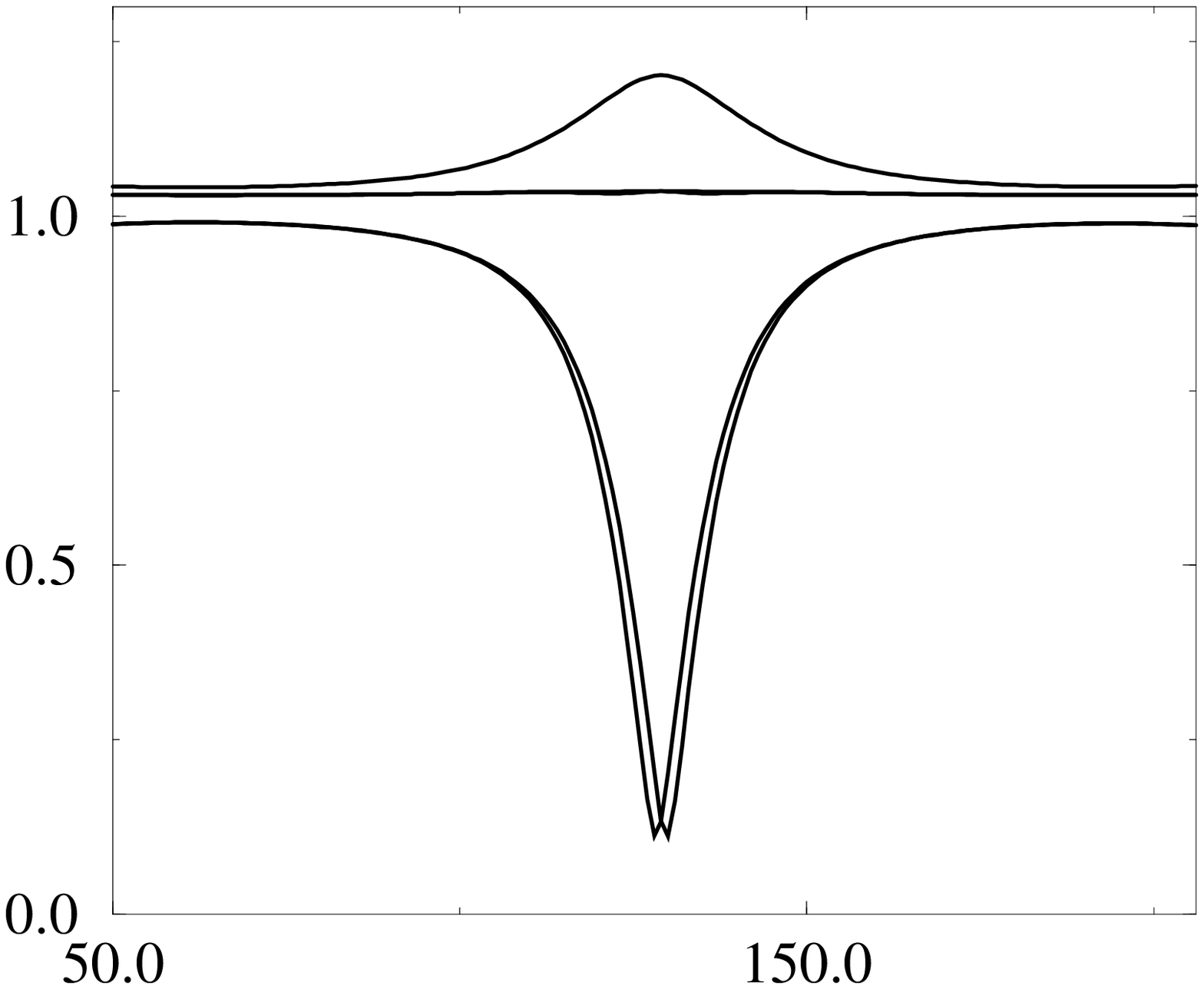}\hspace{2cm}
\epsfxsize=3.5cm\epsfysize=2cm\epsfbox{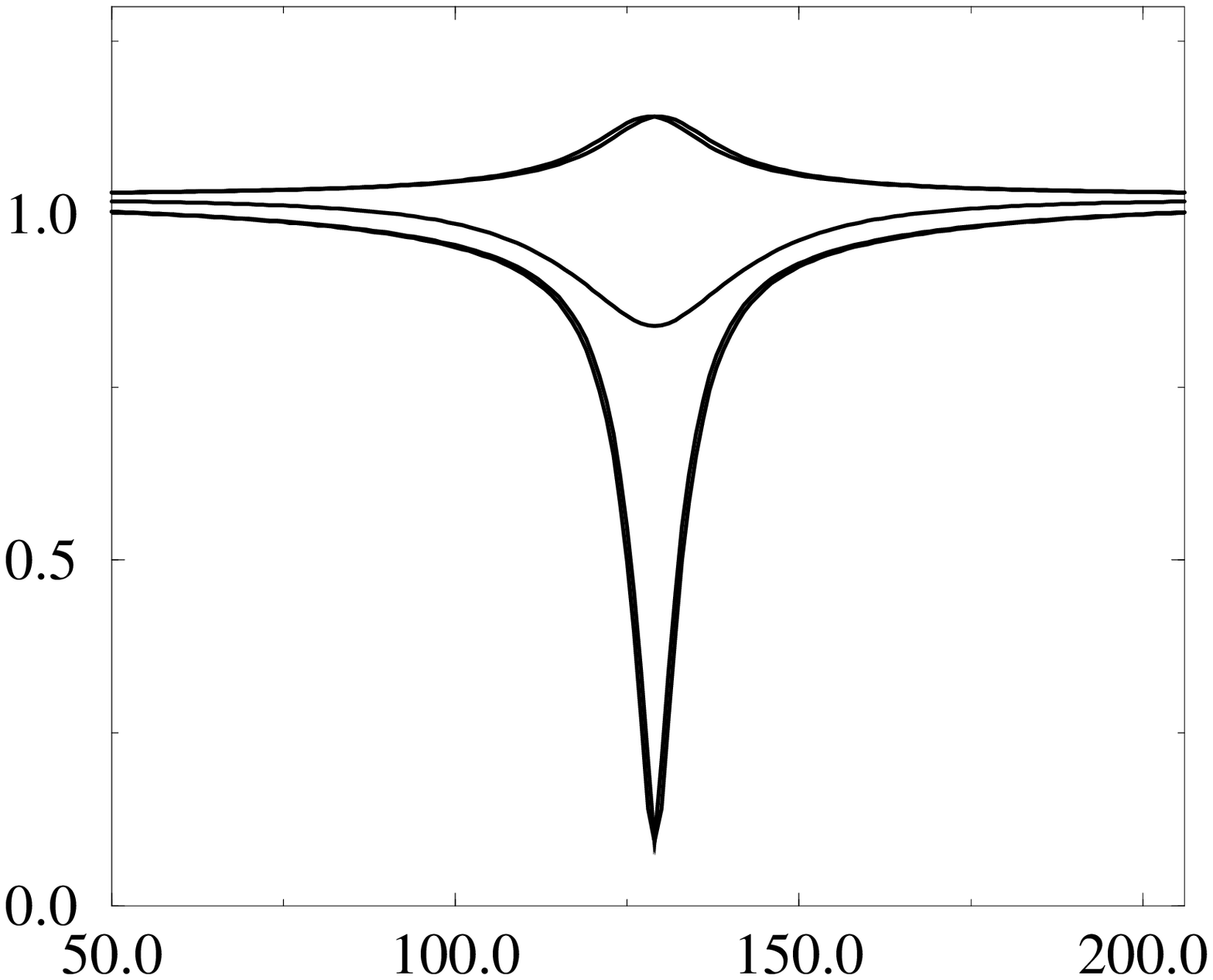}
}

\vspace{0.2cm}

\centerline{
\epsfxsize=4.8cm\epsfbox{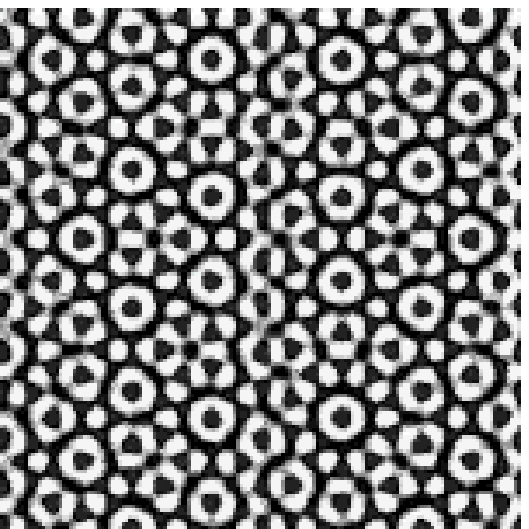}\hspace{0.5cm}
\epsfxsize=4.8cm\epsfbox{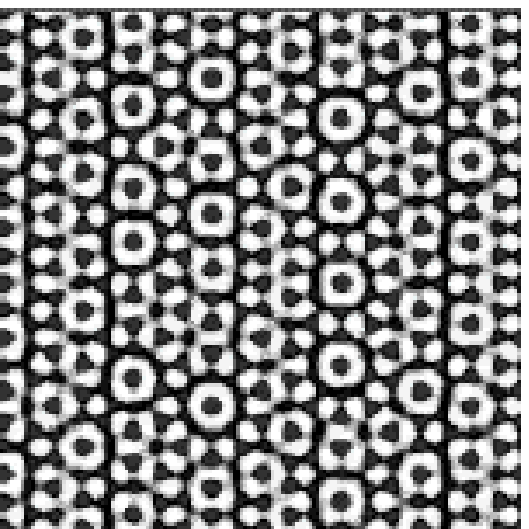}
}
\caption{Decagonal quasipattern with a defect of type a) $H_1$ and b) $H_2$ in 
the center. The top panels show a cross section of the five amplitudes 
in the x-direction. Two of them vanish at the core of the defect. Of the 
remaining three, two are equal, as expected from Eqs. (\ref{eq.H1amp}), 
(\ref{eq.H2amp}). In the bottom panels a reconstruction of the physical field 
$\psi$ is shown. At the core of the defects, the one-dimensional quasipatterns 
$H_1$ and $H_2$ can be observed (compare with Figs. \ref{fig.solsn5}d,e).
\label{fig.defsn10}}
\end{figure}

The decagonal case appears to be the most interesting one in terms of the defect dynamics.
As in the case of defects in hexagon patterns, the (quartic) resonance term
requires that in a defect two amplitudes vanish. In contrast to the hexagonal
or dodecagonal case there are, however,
 two qualitatively different defect types. In one of them the vanishing Fourier modes are rotated 
by $2\pi/5$  with respect to each other while in the other they are rotated by 
$2\times 2\pi/5$. While in the former case 
the core exhibits the quasipattern $H_2$, it is the quasipattern $H_1$ that 
appears in 
the latter case (cf. Fig. \ref{fig.solsn5}d,e). Examples of the two cases are 
shown in Fig. \ref{fig.defsn10}a,b.
Overall, there are $(2\times5)\times(2\times4)/2=40$ different defects.

The same arguments employed in the hexagonal case \cite{Ts96} suggest that 
the interaction between defects is richer in the decagonal quasipattern than 
in hexagons or in the other quasipatterns investigated in the present paper. 
In hexagonal or dodecagonal patterns two penta-hepta defect pairs always share 
a mode that vanishes in both pairs and the charge product $N$ given by 
(\ref{e:Ndef}) never vanishes, implying that the defects always interact 
strongly through the phase. In the decagonal 
quasipattern, however, the analogously defined $N$ (with $n=5$) 
can also take on the value 0, since two pairs of defects need not 
share a mode with vanishing 
amplitude. The 
above argument therefore suggests that in this case the interaction is 
very weak. We have confirmed this expectation by
numerical simulations of the Ginzburg-Landau equations (\ref{eq.GLn10}).
While defect pairs that share a vanishing amplitude ($N \ne 0$)
attract or repel each other quite strongly, those with $N=0$ move so slowly that
our preliminary simulations were not able to identify even whether their interaction
is attractive or repulsive. 

A separation into weak and strong interaction also arises
 in the dodecagonal quasipattern. There, however, the division between 
strong and weak is parallel to the division into the two sublattices.  
In the decagonal case, however, the resonance term involves all modes and the quasipattern
cannot be viewed as the combination of two separate sub-lattices. Correspondingly, 
a defect of one type can change into a defect of any other type through collisions
with suitable other defects. 
We expect that the exceedingly weak 
interaction of defect pairs with $N=0$ may slow down the 
evolution from disordered or random initial conditions towards an
ordered quasipattern. 

\section{Conclusions}

In this paper we have addressed the stability of various types of quasipatterns
with respect to long-wave sideband instabilities. For quasipatterns with
octagonal, decagonal, and dodecagonal rotational symmetry we have derived from the corresponding
Ginzburg-Landau equations long-wave equations for the two phase and 
the two phason modes. Their stability analysis yields the long-wave
stability properties of these patterns. 

For the octagonal and the decagonal 
quasipatterns the phase and the phason modes are
coupled and one cannot clearly separate the instabilities in those of 
the phases and those of the phasons. In these cases our numerical simulations
suggest that the nonlinear behavior
arising from the instabilities is the same as that observed in the 
usual Eckhaus instability. Thus, the instabilities do not saturate within 
the long-wave equations and lead to the creation of defect pairs which
subsequently annihilate each other yielding a stable quasipattern
with slightly modified wavevectors.  Interestingly, however, in the
dodecagonal quasipatterns the phase and phason equations decouple and there are 
parameter regimes in which the quasipatterns first becomes unstable with 
respect to phason modes rather than phase modes. The 
ensuing dynamics appear to be somewhat different than for the usual 
phase modes. This question has, however, not been 
pursued in detail in this paper.

The interaction between defects of the decagonal quasipattern can be 
extremely weak for certain combinations of defects. We expect that this will
have a strong influence on the evolution from disordered initial conditions
to the regular quasipattern. The investigation of the evolution can most likely 
not be addressed within the Ginzburg-Landau equations (\ref{eq.GLn10}) since
the ordering dynamics will also entail a spreading of the modes in 
Fourier space along the critical circle (cf. the dynamics found
in hexagon patterns with rotation \cite{EcRi00}). This will necessitate the use 
of suitable equations of the Swift-Hohenberg type.
  
We gratefully acknowledge discussions with J. Vi\~nals, L. Kramer and M. Silber. 
This research was supported by NASA (NAG3-2113), the Engineering 
Research Program of the Office of Basic Energy Sciences at 
the Department of Energy (DE-FG02-92ER14303) and 
a grant from NSF (DMS 9804673).


\end{document}